\documentclass[a4paper,11pt]{article}
\pdfoutput=1 

\usepackage{jheppub} 
\usepackage[utf8]{inputenc}

\usepackage{setspace}

\usepackage{physics}
\usepackage[obeyFinal]{todonotes}

\def\beq{\begin{eqnarray}}\def\eeq{\end{eqnarray}}
\def\be{\begin{equation}}\def\ee{\end{equation}}

\def\g{\gamma}

\def\g{\gamma}

\def\mes[#1]{d^{3}{#1}}

\def\del{\partial}

\newcommand{\half}{\frac{1}{2}}
\newcommand{\quarter}{\frac{1}{4}}

\def\del{\partial}

\reversemarginpar

\title{Extremal and Near-Extremal  Black Holes and Near- $CFT_1$}

 \author[a]{Upamanyu Moitra,}
 \author[a]{Sandip P. Trivedi,}
 \author[a]{V. Vishal}
 \affiliation[a]{\it Department of Theoretical Physics,
 Tata Institute of Fundamental Research,\\  Colaba, Mumbai, India, 400005\\}

\emailAdd{upamanyu@theory.tifr.res.in}
\emailAdd{sandip@theory.tifr.res.in}
\emailAdd{vishal@theory.tifr.res.in}

\vspace{1cm}

\abstract{We study the behaviour of extremal and  near-extremal black holes at low energies and low temperatures and find that it   can be understood from the near-horizon $AdS_2$ region. Our analysis includes  charged matter and also goes beyond  the $S$-wave approximation. We find that  the leading behaviour at low energies arises from  a mode linked to time reparametrisations and from  phase modes  arising from gauge fields. At somewhat higher energies, additional modes arising from higher partial waves can also be cumulatively significant. These  results can be applied quite generally  to cases where an $AdS_2\times S^d$ near-horizon geometry arises, including black holes in asymptotically $AdS$ and flat space-times. 
}

\preprint{\parbox{3cm}{TIFR/TH/18-28}}

\begin{document}
\maketitle
\flushbottom

\section{Introduction }
The study of extremal and near-extremal black holes and black branes has proved to be a rich and rewarding area of investigation. 
The  low energy and  small temperature response in these systems is expected to arise from the near-horizon region where the gravitational red shift is big.
For several brane configurations it is known that the near-horizon region contains an $AdS_n, n>2$, factor and the resulting low energy dynamics is described by a conformal field theory (CFT) living on the boundary of this region. The CFT plays the role of an effective theory governing the interactions of the black brane with the outside at low energies.  More remarkably,  there is good evidence now that the CFT is often exactly equivalent to the  gravity theory in the interior region described by the  $AdS_n$ geometry. 

In contrast to black branes,  which are extended, the situation vis-\`{a}-vis black holes has been less clear. The near-horizon region for extremal black holes often contain an $AdS_2$ factor. It  is now known that the $AdS_2$ geometry can be used to correctly compute the ground state degeneracy of  extremal black holes to remarkable precision, \cite{Sen:2008yk,Sen:2008vm}.  However, it has also been long suspected that for probing low energy excitations above extremality the $AdS_2$ geometry does not suffice and corrections would need to be incorporated. For example,  it has been  shown  that the back reaction to a probe in $AdS_2$ would diverge rendering the $AdS_2$ unstable to any perturbation, \cite{Maldacena:1998uz}. The underlying physical reason  is  that all the directions transverse to the $AdS_2$ are finite for a black hole and  this simply does not allow the perturbations enough room to dissipate. 

What has been less clear, until recently, is  how to incorporate  the corrections to the $AdS_2$ geometry  in a systematic way in analysing the low energy response. Interestingly, progress in answering this question has  come from the study of quantum mechanical models in condensed matter physics, pioneered by Sachdev and Ye, \cite{Sachdev:1992fk}, and Kitaev, \cite{Kitaev-talks:2015}, and extended in subsequent papers, \cite{Maldacena:2016hyu,Witten:2016iux,Gross:2016kjj,Gurau:2016lzk,Klebanov:2016xxf,Gu:2016oyy,Berkooz:2016cvq,Jevicki:2016bwu,Fu:2016vas,Garcia-Garcia:2016mno,Davison:2016ngz,Krishnan:2016bvg,Li:2017hdt,Eberlein:2017wah,Choudhury:2017tax,Sonner:2017hxc,Yoon:2017nig,Bulycheva:2017uqj,Gross:2017vhb,Murugan:2017eto,Krishnan:2017lra,Bhattacharya:2017vaz,Forste:2017apw,Goel:2018ubv,Chang:2018sve,Blommaert:2018oro,Nosaka:2018iat,Khveshchenko:2017mvj} (we shall refer to the class of such models as SYK models below). It was found that an approximate  conformal theory does arise in the IR in these models, but the breaking of conformal invariance cannot be neglected even at very low energies. This breaking is important in giving a finite action for a mode tied to the breaking of time reparametrisation invariance, and   this mode is in fact responsible for the thermodynamics at low temperatures and    dominates the low energy response of the system.
 
Remarkably, very similar behaviour was found to arise in a theory of two dimensional gravity coupled to a scalar called the dilaton - the Jackiw--Teitelboim (JT)  model, \cite{JACKIW1985343,Teitelboim:1983ux}. In this model the geometry is identically $AdS_2$, with $SL(2,R)$ isometries,  but  the varying dilaton breaks the $SL(2,R)$ symmetry \cite{Maldacena:2016upp}. The fluctuations of the boundary of $AdS_2$ space then acquire a finite action which is of the same form as that in the SYK model, involving the Schwarzian derivative of the time reparametrisation of the boundary. 

The study of SYK models was extended to  models with complex fermions  with conserved charge in \cite{Fu:2016vas,Davison:2016ngz}
It was found that in this case an additional mode arises corresponding to the phase  degree of freedom conjugate to the charge. The breaking of conformal invariance is important for this mode as well, and its  action is also suppressed by the scale at which the conformal invariance is broken. 

The purpose of this paper is to investigate how general are the lessons learnt  from the study of the SYK models for gravitational systems.  The investigation we report on is a continuation of the analysis begun in \cite{Nayak:2018qej} in which a simple system, the Reissner-Nordstr\"{o}m black hole in $AdS_4$, was analysed close to extremality in the $S$-wave sector and its dynamics at low-energies was found to agree with the JT model. More precisely, the low energy response  of the  extremal black hole or near-extremal black hole was found  to arise from the near-horizon region and shown to be well approximated by the JT model. The role of the dilaton in this model  is played by the radius of the $S^2$ transverse to the $AdS_2$ directions and  it was found that the $AdS_2$ region  can {\it in effect} be taken to have a boundary, located at a  fixed value of this radius.  Fluctuations of the  boundary, due to fluctuations in the radius,  then  gives  rise to the time reparametrisation 
 mode mentioned above which dominates the low energy behaviour. 
 
This paper extends the analysis in \cite{Nayak:2018qej} in important ways. By considering  charged matter we  show that  an additional phase mode is generically present with an action of the form given in \cite{Davison:2016ngz} which is suppressed by the scale characterising departures from the near-horizon $AdS_2$ geometry. By studying the correlation functions of probe scalar fields we find  that the exchange of this phase mode  along with the  time reparametrisation mode  correctly gives rise to the low-energy response of the black hole. 
Some of the key formulas containing the time reparametrisation and phase modes along with their coupling to matter fields (eqs. \eqref{modeact}, \eqref{scoup1}, \eqref{scoup2}) and their finite temp counter-parts (eqs. \eqref{actschwa}, \eqref{phasetemp}, \eqref{scoup1t}, \eqref{scoup2t}) are contained in section \ref{bgf}.

We also go beyond the $S$-wave sector. The resulting dynamics can be conveniently analysed in the  two dimensional theory obtained by carrying out a KK reduction over the $S^2$. We show that additional phase modes arise from massless gauge fields which are tied to  isometries of the internal space. In addition, massive fields  also arise from the KK reduction. The effects of these massive states in correlation functions of probe fields can be calculated in $AdS_2$ space with  the  breaking of conformal invariance being neglected. While the $S$-wave sector and additional massless gauge fields dominate at low energies, the cumulative effect of the massive states, which arise from the higher partial waves, can also be   significant at somewhat higher energies.

The lessons we learn in asymptotic $AdS_4$ space can be generalised to other situations, including higher dimensions and asymptotically flat space. 
This allows us to formulate a general method for calculating the low energy behaviour of extremal black holes whose near-horizon geometry is of the form $AdS_2\times S^d$. 

This paper is structured as follows. Having introduced the charged black hole system in $AdS_4$ and reviewed some background  in section \ref{RNBH},  we then analyse the effects of an intermediate scalar  being exchanged in correlation functions in section \ref{scalarsection}. The effects of a probe gauge field, not excited in the background, are studied    in section \ref{probeu1} and  found  to be governed, at leading order, by  the dynamics of a phase mode. In section \ref{bgf} we study a  probe charged scalar field in the presence of the gravity and a background gauge field. These results are put together in section \ref{bgf} and \ref{nads} which discuss how to calculate the response of an extremal  black hole at low energies in general  and extends this analysis  to other situations, including higher dimensions, black holes in asymptotically flat space and multiply charged black holes. 
 
Appendices \ref{probe2du1}, \ref{4dbgf} and \ref{2dbgf} contain useful supplementary details. 
 
It is worth drawing attention to some important references before  we proceed further. See \cite{Castro:2008ms,Castro:2014ima,Jensen:2016pah,Almheiri:2014cka,Maldacena:2016upp,Engelsoy:2016xyb,Almheiri:2016fws,Cvetic:2016eiv,Mandal:2017thl,Gross:2017hcz,Forste:2017kwy,Das:2017pif,Kyono:2017pxs,Eling:2017txo,Dartois:2017xoe,Krishnan:2017txw,Dubovsky:2017cnj,Taylor:2017dly,Grumiller:2017qao,deBoer:2017xdk,Jian:2017tzg,Cai:2017nwk,Halmagyi:2017leq,Kitaev:2017awl,Das:2017hrt,Das:2017wae,Haehl:2017pak,Mertens:2018fds,Nayak:2018qej,Castro:2018ffi,Gaikwad:2018dfc,Lam:2018pvp,Li:2018omr,Harlow:2018tqv,Kolekar:2018sba} for some recent discussion on $AdS_2/CFT_1$ correspondence and extremal black holes.
 
\section{Spherically Symmetric Reissner-Nordstr\"{o}m Black Holes}
\label{RNBH}

Let us describe the geometry of a charged black hole in asymptotically $AdS_4$ space time which arises in a theory of  Einstein gravity with a negative cosmological constant coupled to a Maxwell field. In this paper we take the black hole to be electrically charged. The action is given by
\begin{equation}
\label{act1}
S={1\over 16 \pi G_N}\int d^4 x\sqrt{-g}\,\big(R-2\Lambda\big)\,-\,\frac{1}{4G_N}\int d^4 x \sqrt{-g}\,F_{\mu\nu}F^{\mu\nu}.
\end{equation}
The spherical Reissner-Nordstr\"{o}m black hole solution is given by
\begin{align}
ds^2&=-a(r)^2\,dt^2\,+\,a(r)^{-2}\,dr^2\,+\,b(r)^2\,(d\theta^2\,+\,\text{sin}^2\theta \,d\varphi^2),\nonumber\\
a(r)^2&=1-\frac{2G_NM}{r}+\frac{4\pi Q^2}{r^2}+\frac{r^2}{L^2},\quad b(r)^2=r^2,\label{RNAdS4II}\\
F_{rt}&=\frac{Q}{r^2}.\label{Qe}
\end{align}
Here $M$ and $Q$ are the mass and charge of the black hole, and  $L$ is the AdS$_4$ radius,
\begin{equation}
\label{valL}
L=\sqrt{\frac{3}{|\Lambda|}}.
\end{equation}
For an extremal black hole, the charge and mass get related as
\begin{align}
Q^2 &= \frac1{4\pi} \pqty{r_h^2+\frac{3r_h^4}{L^2}},\label{Qext}\\
M_{ext}&= \frac{r_h}{G_N} \pqty{1+\frac{2r_h^2}{L^2}}\label{Mext}.
\end{align}
With these values of $M$ and $Q$, eq.\eqref{RNAdS4II} becomes
\begin{align} 
\label{axt}
a^2(r) = \frac{ (r-r_h)^2 }{ r^2 L^2 } \, \pqty{ L^2 + 3 r_h^2 + 2 r r_h + r^2 }.
\end{align}
In the first few sections of this paper we will primarily be interested in a big black hole whose   horizon size is much bigger than the $AdS$ radius, 
\begin{equation}
\label{bigb}
r_h \gg L.
\end{equation}
For such big black holes, the charge and mass at extremality can be approximated as 
\begin{align}
Q^2 & \simeq \frac1{4\pi} \pqty{\frac{3r_h^4}{L^2}},\label{Qext2}\\
M_{ext}&\simeq\pqty{\frac{2r_h^3}{G_N L^2}}\label{Mext2}.
\end{align}
Upto $O({r-r_h\over r_h})$ corrections, the metric in the near-horizon region of the black hole is given by
\begin{equation}
\label{nearhor}
ds^2= \bqty{ -\frac{(r-r_h)^2}{L_2^2}dt^2\,+\,\frac{L_2^2}{(r-r_h)^2}dr^2\,+\,r_h^2\,(d\theta^2\,+\,\text{sin}^2\theta \,d\varphi^2) }.
\end{equation}
We see that the near horizon metric is of the form $AdS_2\times S^2$, where  $AdS_2$ and $S^2$ radii are given by
\begin{align}
R_{AdS_2}&=L_2\simeq\frac{L }{\sqrt{6}},\label{ads2rad}\\
R_{S^2}&=r_h.\label{S2rad}
\end{align}
The metric eq.(\ref{nearhor}) is a good approximation when 
\begin{equation}
\label{condaar}
{r-r_h\over r_h}\ll 1.
\end{equation}

The extremal black hole corresponds to a state at chemical potential $\mu$ in the boundary $CFT_3$ dual to the $AdS_4$, with 
\be
\label{chempot}
\mu \sim {r_h\over L^2}.
\ee 

The entropy of the extremal black hole is 
\be
\label{exent}
S_{ext}={\pi r_h^2\over G_N}.
\ee

A near-extremal black hole has a temperature $T\ll \mu$. 
Its entropy is given by,
\be
\label{nexen}
S-S_{ext}= \frac{4\pi^2}{G_N}\,L_2^2\,r_h\, T,
\ee
and the free energy is
\begin{equation}
\label{free}
F=-\frac{1}{\beta}\,\frac{\pi r_h^2}{G_N}-\frac{2\pi^2}{G_N}L_2^2\,r_h\,T^2.
\end{equation}
We see that the specific heat is linear near extremality. 

It is known that the thermodynamical description breaks down very close to extremality; this has been often found to agree with the existence of a gap in the system. 
The break down occurs when the specific heat is order unity, leading to the condition, 
\be
\label{Tg}
T_{gap}\sim\frac{G_N}{L_2^2 r_h}.
\ee

We are interested in the low energy behaviour of the system which corresponds to being at a  temperature $T\ll \mu$ and to studying the response at energies\footnote{More generally, see eq.(\ref{condff}).}  $\omega \ll \mu$. Using eq.\eqref{chempot} these conditions become,
\be
\label{lowT}
T\ll {r_h\over L_2^2}
\ee
and 
\be
\label{condo2}
\omega \ll \frac{r_h}{L_2^2}.
\ee
We will also take $T\gg T_{gap}$.

Corrections to the  $AdS_2\times S^2$  geometry, especially the  change in the radius of the $S^2$, are important  to keep in analysing the low energy behaviour, \cite{Maldacena:2016upp,Nayak:2018qej}. The response of the system can be understood as arising from the region obtained  by cutting off the near horizon geometry 
along a boundary which corresponds to a constant value of the dilaton. This boundary will lie in the asymptotic $AdS_2$ region. Working in Poincar\'{e} coordinates,
\be
\label{zdef}
z={L_2^2 \over (r-r_h)},
\ee
the metric of $AdS_2$ can be written as
\be
\label{metads2}
ds^2=\frac{L_2^2}{z^2}\pqty{dt^2+dz^2}.
\ee
The boundary lies in the region where 
\be
\label{asb}
\omega z\ll m L_2\sim O(1).
\ee
From eq.(\ref{zdef}) we see that this  implies 
\be
\label{asb2}
\omega L_2^2 \ll (r-r_h)
\ee
which is consistent with eq.(\ref{condaar}) since  eq.(\ref{condo2}) is being met. 

It is also useful to briefly discuss the asymptotically flat case (which corresponds to setting $\Lambda=0$ in eq.\eqref{act1}). The behaviour of small black holes, with $r_h \ll L$ in $AdS_4$ is similar. The solution  for the extremal RN black hole in asymptotically flat space is  given by 
\begin{equation}
ds^2=\pqty{1-\frac{\sqrt{4\pi}\,Q}{r}}^2dt^2+\pqty{1-\frac{\sqrt{4\pi}\,Q}{r}}^{-2}dr^2+r^2\,d\Omega_2^2\label{flatext}.
\end{equation}
The radius of the $S^2$ at the horizon is $r_h=\sqrt{4\pi}\,Q$. 
The near horizon geometry is $AdS_2\times S^2$ with both the radius of the $S^2$ and $AdS_2$ being $r_h$, i.e., 
\be
\label{asl2}
L_2=r_h.
\ee 
This is in contrast with big black holes in $AdS_4$; these have  a value of  $L_2$   given in eq.\eqref{ads2rad} which  does not depend on the charge $Q$. 

From the field strength, working in the gauge $A_r=0$, we get that
\be
\label{chempa}
A_t=\frac{Q}{r_h}-\frac{Q}{r}.
\ee
The chemical potential of the black hole $\mu$  is the value of $A_t$ at  $r=\infty$; we see that $\mu =\frac{Q}{r_h}$ and does not grow with the charge $Q$ since the radius of the black hole is also growing linearly with $Q$\footnote{Note from eq.\eqref{act1} that in our units $F_{\mu\nu}$ is of dimension $1$ so that the chemical potential is dimensionless.}. This is in contrast to big black holes in $AdS_4$ which we have discussed above for which $Q\sim \frac{r_h^2}{L}$ and $\mu \sim \frac{r_h}{L^2}$, so that the chemical potential does grow with the charge. 

Despite these differences some of the key features of the analysis for big black holes in $AdS_4$ can be easily extended and apply to black holes in asymptotically flat space as well. Corrections to the near horizon geometry are small as long as eq.(\ref{condaar}) is met.  It can be easily seen that the condition for low energy and small temperature,  so that the response arises from the near horizon region,   still takes the form, 
 eq.(\ref{lowT}), eq.(\ref{condo2}). From eq.(\ref{asl2}) these become, 
\begin{align}
T\ll \frac{1}{r_h}\label{flatT},\\
\omega\ll \frac{1}{r_h}\label{flatom}.
\end{align}

A similar  analysis, for  both  black holes in asymptotically $AdS$ and flat space, also applies in higher dimensions.

\section{Four Point Function with Scalar Exchange in the Bulk}
\label{scalarsection}
In this section we study a probe scalar $\psi$ which is coupled to another scalar $\phi$.

The  exchange of this $\phi$ particle then gives rise to a four-point function for $\psi$. We will analyse the resulting four-point function in a partial wave expansion. Our conclusions  will also generalise to  contributions to the four-point function due to graviton or gauge field exchange.

We work in Euclidean space in this section. 
The  matter action is
\begin{align}
S_M={1\over 2G_N}\bigg( & \int d^4x \sqrt{g} \pqty{  \del_{\mu}\psi\,\del^{\mu}\psi+m_{\psi}^2\psi^2}\,+\,\int d^4x \sqrt{g} \pqty{\del_{\mu}\phi\,\del^{\mu}\phi+m_{\phi}^2\phi^2}\nonumber \\ &+\lambda \int d^4x \sqrt{g}\, \phi \psi\psi \bigg).\label{matact}
\end{align}
Here we have coupled the probe $\psi$  by a cubic vertex to the scalar $\phi$. 

At low energies, eq.\eqref{condo2},  the correlation functions for the operator dual to $\psi$ arise from the near-horizon $AdS_2$ region, as discussed in section 3 of \cite{Nayak:2018qej}. A sufficient condition  is 
\be
\label{condff}
\omega \ll m\, {r_h\over L_2}.
\ee
Setting $m \sim O(\frac{1}{L_2})$ gives, eq.\eqref{condo2}.

The metric of $AdS_2\times S^2$ in Poincare coordinates is given by 
\be
\label{poincare}
ds^2={L_2^2\over z^2} \pqty{dt^2+dz^2} + r_h^2 d\Omega^2.
\ee

Expanding $\psi$ in terms of the KK harmonics on the $S^2$ we get,
\be
\label{expkk}
\psi=\sum_{lm} \psi_{lm}(z,t) Y_{lm}(\theta,\phi).
\ee
The fields $\psi_{lm}(z,t)$ are scalars in $AdS_2$ with  mass 
\be
\label{masspsi}
m_{\psi, l}^2=m_\psi^2 + {l(l+1) \over r_h^2}.
\ee
Using the standard dictionary it is easy to see that they correspond to operators with dimension
\be
\label{dimads2}
\Delta^{(l)}_{+}=\half+\sqrt{{1\over 4} +m_{\psi,l}^2L_2^2}.
\ee
Similarly for $\phi$ we get the KK modes $\phi_{lm}(z,t)$ with mass
\be
\label{massphi}
m^2_{\phi,l}=m_\phi^2+ {l(l+1) \over r_h^2}.
\ee

Substituting eq.\eqref{expkk} in the action above eq.\eqref{matact}  gives, 
\begin{align}
\label{sma}
S_{matter} & =  {r_h^2\over G_N} \sum_{lm} \int d^2x\sqrt{g} \,\half\,\Big(|\partial \psi_{lm}|^2 + m_{\psi,l}^2 |\psi_{lm}|^2+ |\partial \phi_{lm}|^2+ m_{\phi,l}^2 |\phi_{lm}|^2\Big) \nonumber \\
&\quad +{ r_h^2\over G_N}  \, \lambda\, 
\pqty{\sum_{l,l'l'';m,m'm''} C^{\{m,m',m''\}}_{\{l,l',l''\}}\int d^2 x \sqrt{g}\, \phi_{l''m''}\psi_{l,m} \psi_{l'm'}},
\end{align}
where we have integrated over the $S^2$ and the metric in the expression is the 2-D metric. In the expression above,

\be
\label{defcl}
C^{\{m,m',m''\}}_{\{l,l',l''\}}=\int d^2\Omega \  Y_{lm}Y_{l'm'}Y_{l''m''}.
\ee

Let us understand the contribution made by the various partial waves in more detail.

The value of $\psi_{lm}$ in the $AdS_2$ bulk can be expressed in terms of the value it takes on the $AdS_2$ boundary located at 
\be
\label{defb}
z=\delta
\ee
 using the bulk to boundary propagator in $AdS_2$,
\be
\label{valpsi}
\psi_{lm}(z,t) =  \int d\omega\,{\hat K}_{\nu_l}(\omega z)\,   e^{-i\omega t}\, \psi_{lm}(\omega).
\ee
Here $\psi_{lm}(\omega)$, the boundary value for frequency $\omega$, corresponds to the source of the dual operator turned on at the $AdS_2$ boundary. 
In eq.(\ref{valpsi}) $\hat K_\nu$ can be expressed in terms of the modified  Bessel function $K_\nu(\omega z)$ as,
\be
\label{defhatk}
{\hat K}_{\nu_l}(\omega z)=\delta^{\Delta^{(l)}_-}\,\frac{\sqrt{\omega z}\,K_{\nu_l}(\omega z)}{\sqrt{\omega \delta}\,K_{\nu_l}(\omega \delta)}
\ee
with 
\be
\label{defnu}
\nu_l=\sqrt{{1\over4}+m_{\psi,l}^2L_2^2},
\ee
and, 
\be
\label{delpm}
\Delta^{(l)}_\pm={1\over 2} \pm \nu_l.
\ee

The bulk Green function for $\phi_{lm}$ satisfies the equation 
\be
\label{gfa}
[\nabla^2 - m_{\phi,l}^2]G_{lm} = {\delta^2(x-x')\over \sqrt{g_2}}.
\ee
In frequency space, this gives
\be
\label{gfb}
\pqty{\partial_z^2 -\ \omega^2 - \frac{m_{\phi,l}^2L_2^2}{z^2}}  G_{lm}(\omega, z, z')= \delta(z-z').
\ee
The resulting on-shell action is then given by 
\begin{align}
S_{onshell} & =  -{\lambda^2\over 2} {r_h^2\over G_N} \sum_{l_i,m_i} \Biggl[ \int \pqty{\prod _{i=1}^{i=4}\, \frac{d\omega_i}{2\pi}\, \psi_{l_im_i}(\omega_i)}\,\delta \pqty{\sum \omega_i} \sum_{l,m} C^{\{m,m_1,m_2\}}_{\{l,l_1,l_2\}} C^{\{-m,m_3,m_4\}}_{\{l,l_3,l_4\}}\nonumber\\
&\int dz\,\frac{L_2^2}{z^2} \int dz'\,\frac{L_2^2}{z'^2}\,\,  {\hat K}_{\nu_{l_1}}(\omega_1z ) {\hat K}_{\nu_{l_2}}(\omega _2 z) {\hat K}_{\nu_{l_3}}(\omega_3 z') {\hat K}_{\nu_{l_4}}(\omega_4z')\, G_{lm}(\omega_1+\omega_2,z,z') \Biggr]\label{4ptfinal}.
\end{align}
This has four powers of the source $\psi_{lm}(\omega)$ and thus is a contribution to the $4$-point function. 

To understand this complicated looking result, it is useful to recall  the scaling dimensions of various fields. 
The quadratic (free) part of the action gives the two-point function for $\psi$, which is given by 
\be
\label{twopsi}
S_{2pt}= {r_h^2 \over G_N} C_1 \sum_{lm}\int d\omega\, \psi_{l,m} \psi_{l,-m} \, \omega^{2\Delta_l},
\ee
where $C_1$ is a constant and $2\Delta_l=\Delta^{(l)}_+-\Delta^{(l)}_-$.
From this it  follows that under the scaling $\omega \rightarrow \lambda \omega$, $\psi_{lm}(\omega)\rightarrow \lambda^{-(\Delta_l+{1\over 2})}$, i.e. $\psi_{lm}(\omega)$ has dimensions $-(\Delta_l+{1\over 2})$. 
It also follows that the operator ${\cal O}_{lm}$ for which $\psi_{lm}$ is the source has dimension $(\Delta_l - {1\over 2})$ so that 
\be
\label{sco}
S_{coupling}= \int d\omega\, \psi_{lm}(\omega){\cal O}_{lm}(\omega) 
\ee
is invariant. 

Now consider a contribution to the four-point function of the form
\be
\label{confp}
S_{4pt}=\prod_{i} \int d\omega_i \,\psi_{l_im_i}(\omega_i) <{\cal O}_{l_1m_1}{\cal O}_{l_2m_2}{\cal O}_{l_3m_3}{\cal O}_{l_4m_4}>.
\ee
By the scaling symmetry of $AdS_2$ we see that 
\be
\label{expf}
<{\cal O}_{l_1m_1}{\cal O}_{l_2m_2}{\cal O}_{l_3m_3}{\cal O}_{l_4m_4}> \sim \omega^{\sum_i \Delta_i - 2}
\ee
where $\Delta_i=\Delta_{l_i}$.
Just from the scaling properties it therefore follows that the operators with higher higher values of $l$, which have bigger anomalous dimensions, eq.(\ref{masspsi}),  will make a smaller contribution at low frequencies. 
By comparison with eq.\eqref{4ptfinal} we see that 
\begin{align}
\label{def4}
<{\cal O}_{l_1m_1}{\cal O}_{l_2m_2}{\cal O}_{l_3m_3}{\cal O}_{l_4m_4}> & =  -{\lambda^2 \over 2} {r_h^2\over G_N} \delta \pqty{\sum_i\omega_i} \Biggl[\sum_l C^{\{m,m_1,m_2\}}_{\{l,l_1l_2\}}C^{\{-m,m_3,m_4\}}_{\{l,l_3l_4\}} \nonumber \\ &\hspace{-85pt} \int dz\,\frac{L_2^2}{z^2} \int dz'\,\frac{L_2^2}{z'^2}\,\, 
 {\hat K}_{\nu_{l_1}}(\omega_1z ) {\hat K}_{\nu_{l_2}}(\omega _2 z) {\hat K}_{\nu_{l_3}}(\omega_3 z') {\hat K}_{\nu_{l_4}}(\omega_4z') G_{lm}(\omega_1+\omega_2,z,z') \Biggr]
\end{align}
The sum over $l$ arises from the contribution of the different intermediate partial waves of the $\phi$ field. If the $\psi_{l_i,m_i}(\omega)$ are comparable for the different $l_i,m_i$ values, the $S$-wave in eq.(\ref{4ptfinal})  will dominate at sufficiently low energies and in this case the sum over $l$ in eq.(\ref{def4}) will only get a contribution from  the $S$-wave for the intermediate field $\phi$. From the behaviour of the Bessel functions $K_\nu$ and the Green function $G_{lm}$  which appear in eq.(\ref{def4}), it is easy to see directly that the four-point function scales with frequency as given in eq.(\ref{expf}).

An important fact to note from eq.(\ref{4ptfinal}), eq.(\ref{def4}) is that the $4$-point function due to the intermediate scalar $\phi$ being  exchanged  is $O(r_h^2)$. In contrast, it was found in \cite{Nayak:2018qej} that $S$ wave contribution due to gravitational effects  goes like $O(r_h^3)$ and is further enhanced by a factor of $r_h$. We will return to this point in section \ref{bgf}.

 \section{Charged Matter Coupled to a Probe  Gauge Field}
 \label{probeu1}
 Next, we consider the behaviour  of a probe gauge field which is not turned on in the background, i.e. with respect to which the black hole is not charged.    The action is  
 \be
 \label{actgf}
 S_{gf}=-{1\over G_N}\int d^4x \sqrt{g}\,\pqty{{1\over 4} F^2 +e\, A_\mu J^\mu},
 \ee
 where $e$ is the charge of the scalar field.
 The current $J^\mu$ is quadratic in the charged matter field. For a charged scalar $\psi$,
 \begin{equation}
  J_{\mu} = i (\psi\, \partial_{\mu} \psi^\dagger-\psi^\dagger \partial_{\mu}\psi).\label{current}
  \end{equation} 
 For concreteness, we are  interested in calculating the four-point function for $\psi$ with an intermediate exchange of the  probe gauge field.
 In particular, we are    interested  in studying the $S$-wave sector of this theory. 
 
 It was noted in \cite{Davison:2016ngz}  that an additional  mode arises in the SYK model with charged matter.  The departure from conformal invariance is important for the dynamics of this mode   whose action is suppressed by the energy scale parametrising this departure. 
 The behaviour of this mode is therefore analogous to the time reparametrisation modes which arise in the $S$-wave sector of gravity.
 
 We will see below that in fact a similar mode   does arise for a $U(1)$ gauge field in the bulk in the $S$-wave sector. The exchange of this mode, which we will refer to as the ``phase mode",
 will dominate the low-energy contribution to the four-point function due to gauge field exchange. Let us note that such a mode was found earlier in $AdS_3$ gravity with a Chern-Simons action for the gauge field in \cite{Gaikwad:2018dfc}.

 The $S$ wave sector can be analysed  in $4$ dimensions directly. We work in the gauge
 $A_t=0$. The source $J^\mu$ is spherically symmetric and therefore only gives rise to a non-trivial value for $A_r$; we can therefore also set $A_\theta, A_\phi=0$. 
 The  equation  of motion is 
 \be
 \label{maxres}
 \partial_\beta(\sqrt{g} F^{ \beta \alpha} )=e\, \sqrt{g} J^{\alpha}
 \ee
 with 
 $J^{\alpha}$ being  nonzero for $\alpha=r,t$. Here  we are using the coordinates in eq.(\ref{RNAdS4II}). Setting $\alpha=r$ gives,
 \be
 \label{solg}
 F^{tr}=e\,\partial_t^{-1} J^r.
 \ee
  The reader should not be alarmed by the factor of $\partial_t^{-1}$. We are  considering time dependent sources here and eq.(\ref{maxres}) can be solved for each frequency, $\omega$,  separately with   $\partial_t^{-1}$  simply being  short hand for a factor of $\frac{1}{\omega}$. Eq.(\ref{solg}) allows for an extra integration constant which is only a function of $r$, this has been set to vanish since we are computing the response to the time dependent source.   Eq.(\ref{solg}) can be solved for $A_r$ to give
  \be
  \label{solar}
  A_r=e\,g_{tt}\,\partial_t^{-2} J_r.
  \ee
  
  Substituting in eq.(\ref{actgf}) gives the on-shell action
  \be
  \label{gfos}
  S_{onshell}=-{1\over G_N} \,4 \pi \,e^2 \int d^2x \sqrt{g_2}\, r^2 \,\pqty{{1\over 2} \,g^{rr}g_{tt}\,J_r \partial_t^{-2} J_r}.
  \ee
   As was noted above,  the current $J_r$ is  quadratic in the charged matter field. 
  Thus eq.(\ref{gfos}) is a contribution to the four-point function for the matter field. In eq.(\ref{gfos}) we have already  integrated over the $S^2$ and the  remaining two dimensional integral  left is over  $r,t$, with the metric $g_{\alpha \beta}$ being the $r-t$ plane metric, and $g_2 $ denoting its determinant. 
  
  At first sight it might seem that the resulting contribution to the four-point function is of $O(r_h^2)$ due to the factor of $r^2$ inside the integral in eq.\eqref{gfos} and not of  $O(r_h^3)$ as in the $S$-wave sector of gravity. However, this conclusion is premature, as we now discuss. 
  
  The charged field $\psi$ we are considering has the action
\begin{align}
S_{matter}&=\frac{1}{G_N}\int d^4x\,\sqrt{g}\,\pqty{|D\psi|^2+m^2|\psi|^2}\label{actpsi},\\
D_{\mu}&=\del_{\mu}-i\,e\,A_{\mu}\label{ddef}.
\end{align}
  with mass $m$.
  
  Now consider the asymptotic $AdS_2$ region. In this region $\psi$ takes the form (see appendix \ref{probe2du1}),
  \be
  \label{formp}
  \psi(z,t)=\psi(t) z^{\Delta_-} + c \,(z^{\Delta_+}-z^{\Delta_-}\,\delta^{\Delta_+-\Delta_-})\int dt'\,\frac{\psi(t')}{|t-t'|^{2\Delta_+}},
  \ee
  where $c$ is a constant, eq.\eqref{cvalue}, and $\Delta_{\pm}$ are given by
  \begin{equation}
  \label{deltapm}
  \Delta_{\pm}=\half\pm\sqrt{\quarter+m^2 L_2^2}.
  \end{equation}
  
  Working in the $z$ coordinate, eq.\eqref{zdef}, the contribution to the integral in eq.(\ref{gfos}), is then given by, 
  \be
  \label{fp2}
  \Delta S_{onshell}=-\frac{r_h^2}{G_N}\,2\pi \,e^2\int dt\,dz \,\sqrt{g}\,g_{tt}\, g^{zz}\, J_z(z,t) \partial_t^{-2} J_z(z,t).
  \ee
  Discarding contact terms it is easy to see from eq.\eqref{jsol} that  
  \be
  \label{jz}
  J_z(t,z)= i\, c\,(\Delta_+-\Delta_-)\int dt'\pqty{\frac{\psi(t)\psi^{\dagger}(t')-\psi(t')\psi^{\dagger}(t)}{|t-t'|^{2\Delta_+}}}.
  \ee
  In particular $J_z$ is independent of $z$. 
  Thus the $z$ dependence in eq.\eqref{fp2}  goes like 
  \be
  \label{4p3}
  \Delta S_{onshell} \sim \int dz {1 \over z^2} \sim \frac{1}{z_c},
  \ee
  where $z_c$ is the value at the boundary of the asymptotic $AdS_2$ region. 
  
  One expects of course that this divergence is cut off once the corrections to the $AdS_2$ geometry become significant. Since this occurs at a scale,
  \be
  \label{sccut}
  z\sim \frac{1}{r_h},
  \ee
 (see eqs.\eqref{condaar} and \eqref{zdef}) we get  by setting $z_c\sim \frac{1}{r_h}$ in eq.(\ref{4p3}) the resulting  four-point function to be of order, 
  \be
  \label{4p4}
  S_{4pt}\sim r_h^3 \int dt \,dt' dt''\pqty{\frac{\psi(t)\psi^{\dagger}(t'')-\psi(t'')\psi^{\dagger}(t)}{|t-t''|^{2\Delta_+}}}\del_t^{-2}\pqty{\frac{\psi(t)\psi^{\dagger}(t')-\psi(t')\psi^{\dagger}(t)}{|t-t'|^{2\Delta_+}}}.
  \ee
  Note that due to the divergence the resulting $4$ point function is in fact enhanced and  of order $r_h^3$, as expected in analogy with the SYK model. 
  In appendix \ref{probe2du1}, we calculate the coefficient  in eq.(\ref{4p4}) precisely. The resulting $4$ point function is then given by 
  \begin{align}
  \label{final4p}
  S_{4pt}&= \pqty{\frac{r_h^2}{G_N}} \,r_h\,2\pi\,c^2\,e^2 (\Delta_+-\Delta_-)^2\nonumber\\
  &\quad\qquad \times\int dt\, dt' dt''\pqty{\frac{\psi(t)\psi^{\dagger}(t'')-\psi(t'')\psi^{\dagger}(t)}{|t-t''|^{2\Delta_+}}}\frac{1}{\del_t^2}\pqty{\frac{\psi(t)\psi^{\dagger}(t')-\psi(t')\psi^{\dagger}(t)}{|t-t'|^{2\Delta_+}}}.
  \end{align}
  
  We have focussed on the asymptotic $AdS_2$ region above. It is easy to see that the interior of the $AdS_2$ region  gives a contribution to the $4$ point function which is $O(r_h^2)$ and is therefore less important. 
  
  \subsection{ The Phase Mode}
  \label{phase}
  The $4$-point function we have obtained above can be understood as arising due to the exchange of a mode with an action suppressed by  $\frac{1}{r_h}$. 
  Working in the $A_t=0$ gauge we have chosen above, take $A_r$ (in the $(t,r)$ coordinate system) to be of the form\footnote{We are grateful to Juan Maldacena for a discussion on this topic.}
  \be
  \label{formar}
  A_r= \chi(r)\, \theta(t)
  \ee
  with 
  \be
  \label{prof}
  \chi(r)=\frac{r_h}{\sqrt{g}\,g^{tt} g^{rr}}.
  \ee
  Here the metric components refer to the extremal geometry given in eq.\eqref{RNAdS4II}\, and $\sqrt{g}=r^2$ is the determinant of the four dimensional metric. 
  
  With the form of $A_r$ in eq.(\ref{prof}), the Maxwell action
  \begin{equation}
  \label{max}
  -\frac{1}{4G_N}\int d^4x\,\sqrt{g}\,F^2
  \end{equation}
gives
  \be
  \label{actp}
  S_{phase}=-\pqty{\frac{r_h^2}{G_N}}\,2\pi \int d^2x\,\frac{1}{\sqrt{g}\,g^{rr}g^{tt}}\,(\dot\theta)^2,
  \ee
  where we have integrated over the $S^2$.
  The upper limit of the $r$ integral is the $AdS_4$ boundary. The lower limit of the integral can be taken to be the $AdS_2$ horizon (up to corrections subleading in the frequency $\omega$, as discussed in the appendix \ref{probe2du1}). This gives, 
  \be
  \label{actp2}
  S_{phase}=  -\pqty{\frac{r_h^2}{G_N}}\,2\pi\,{1\over r_h} \int dt \,(\dot\theta)^2.
  \ee
   We will see below that $\theta(t)$ plays the role of  the phase mode mentioned above. For now we note that its action is indeed suppressed by the scale $r_h$ which is related to the  energy scale 
   $\mu \sim \frac{r_h}{L^2}$ at which departures from the $AdS_2$ geometry become significant.

  Let us see how this  mode enters in  the thermodynamic properties. Consider  a black hole in $AdS_2$ which now also  carries  charge $q$ with respect to this probe field. 
  As discussed in appendix \ref{2dbgf}, to calculate the partition function for the grand canonical ensemble, the action has an extra term compared to the canonical ensemble \cite{Hawking:1995ap}.
    This results in an extra contribution in the free energy of the grand canonical ensemble: 
  \be
  \label{extrac}
  \Delta F= -\frac{4\pi}{G_N}{q^2 \over r_h}.
  \ee
  
  This additional contribution is given by twice the contribution of the phase mode $S_{phase}$ if we identify 
  \be
  \label{idq}
  q=r_h\,\dot\theta.
  \ee
  Note that  eq.(\ref{idq}) solves the equation of motion for $\theta$ (we are considering the system without sources here) :
  \be
  \label{eom}
  \ddot{\theta}=0.
  \ee

  Next let us see how the exchange of this mode gives rise to the $O(r_h^3)$ contribution to the $4$ point function we obtained above. 
  The charged matter field $\psi$, eq.\eqref{actpsi},  has a two-point function which arises from an action given by 
  \be
  \label{twopt}
  S_{2pt}= -\pqty{\frac{r_h^2}{G_N}}\,4\pi \,c \, (\Delta_+ - \Delta_-) \int dt\, dt' \,\frac{\psi^\dagger (t) \psi(t')}{|t-t'|^{2\Delta_+}}.
  \ee
  The action is evaluated at the boundary of the near $AdS_2$ region, with $\psi(t)$ being the boundary value for $\psi$. The field $\theta(t)$ is to be identified  
  with the phase of $\psi$. This results in a coupling (at linear order in $\theta)$:
  \be
  \label{2pt2}
  S_c=\pqty{\frac{r_h^2}{G_N}}\,4\pi \,c\,(\Delta_+-\Delta_-)\,e\,i\int dt\, dt' \,\frac{\psi^\dagger (t) \psi(t')}{|t-t'|^{2\Delta_+}}\,\pqty{\theta(t)-\theta(t')}.
  \ee

  Integrating out $\theta$ using eq.\eqref{actp2} we now get a $4$-pt function which agrees with eq.\eqref{FHact}.
  Higher point functions, to leading order in ${r_h^2\over G_N}$, arise from successive tree level exchanges of the phase mode, and can calculated in a similar fashion. 
  
  \subsection{Additional comments}
  \label{adcom}
  It is worth noting that even though the radial integral involved in obtaining the full $O(r_h^3)$  term got its contribution from the region extending beyond the near-horizon geometry,
  the final result for the $4$-point function could be expressed in terms of the value  for $\psi$ at the boundary of the near-horizon region, eq.\eqref{formp}. This result is expected on physical grounds since the low-energy response should arise from the near-horizon region. 
  
  It is also worth mentioning that the phase mode does not correspond to a gauge transformation in the $AdS_2$ region, since $A_r$ in eq.\eqref{formar} gives rise to a non-trivial field strength when $\theta$ is time dependent. What the mode does capture is both the charge dependence in the grand canonical ensemble and the Coulomb dressing effects for charged perturbations in the $S$-wave. Roughly,  this mode can be thought of as a Wilson line  (smeared on the $S^2$) in the radial direction extending from the horizon to the boundary of $AdS_2$.

  We have concentrated on the $S$-wave sector so far. The  higher partial waves of the probe gauge field will give rise to massive gauge fields in the two dimensional theory obtained by dimensional reduction to $AdS_2$. In addition, scalars will also arise from the KK reduction to two dimensions. These extra states will then also contribute   as intermediate channels in the four-point function.  To leading order their contribution   can be calculated in the $AdS_2$ geometry itself,  neglecting any departures, in a manner analogous to the effect of the intermediate  scalar in section \ref{scalarsection}. 
  
Expressing the $4$ pt. function in Fourier space,
\be
\label{4pf}
S_{4pt}=\pqty{\prod_{i=1}^{4} \int \frac{d\omega_i}{2\pi}} \psi(\omega_1) \psi(\omega_2) \psi^\dagger(\omega_3) \psi^\dagger(\omega_4)<O(\omega_1) O(\omega_2) O(\omega_3)^\dagger O(\omega_4)^\dagger>,
\ee
it is easy to see  from eq.\eqref{final4p} that the $S$-wave contribution to $<O(\omega_1)O(\omega_2)O(\omega_3)^\dagger O(\omega_4)^{\dagger}>$ will scale with frequency as\footnote{The engineering dimensions are accounted for by suitable powers of $L_2$.}
\be
\label{scf}
<O(\omega_1) O(\omega_2) O(\omega_3)^\dagger O(\omega_4)^\dagger>\sim e^2\pqty{{r_h^2\over G_N}} r_h \,\omega^{4 \Delta -3}.
\ee
This might seem a bit surprising at first, since the anomalous dimension in $AdS_2$ for $O$ would result in a scaling, see eq. (\ref{expf}), 
$<OOO^\dagger O^\dagger>\sim \omega^{4 \Delta -2}$.

The situation here is similar to what happens with the Schwarzian action for time reparametrisation modes. 
The action for the phase mode, eq.\eqref{actp2},  breaks the scaling symmetry of $AdS_2$ under which $t\rightarrow \lambda t$ (with $\theta(t) \rightarrow \theta(\lambda t))$.
The breaking is due to the factor of $\frac{1}{r_h}$ which suppresses the action; its presence we saw above  is  tied to the fact that the action arises not in the near $AdS_2$ region but in the region which interpolates from the asymptotic $AdS_2$ geometry to the $AdS_4$ boundary. 

A convenient way of book keeping is to assign the factor of $r_h$ which suppresses the action in eq.\eqref{actp2} a transformation law $r_h \rightarrow { r_h \over \lambda}$ to keep track of this breaking. It is important to bear in mind that we do this while keeping the prefactor ${r_h^2\over G_N}$ in eq.\eqref{actp2} invariant under the rescaling. The point being that ${r_h^2\over G_N}\sim S_{ext}$ is the extremal entropy and therefore a measure of the degrees of freedom in the IR $AdS_2$, while the extra factor of $\frac{1}{r_h}$ is due to the breaking of conformal invariance. It is easy to  see that eq.\eqref{actp2} is invariant with this extra rescaling of $r_h$. It also then follows that the $4$ pt. function scales as given in eq.(\ref{scf}) since it has an extra factor of $r_h$ in front, eq.\eqref{final4p}.

   Consider now higher partial waves of the $\psi$ field. For suitable values of their angular momentum the  contribution of the $S$ wave phase mode could still arise. For any mode
    $\psi_{lm}$  with angular momentum $l$, the two-point function can be written in the form, eq.\eqref{twopt} with $\Delta_l$ being the anomalous dimension for the mode. This would then lead to a coupling with the phase mode  analogous to eq.\eqref{2pt2},  resulting in a $4$ pt function due to the exchange of the phase mode. In frequency space this would lead to
   \be
   \label{fm2}
   <O_l(\omega_1)O_l(\omega_2)O_{l'}(\omega_3)^\dagger O_{l'}(\omega_4)^\dagger> \sim e^2\, \pqty{{r_h^2\over G_N}}\, r_h\, \omega^{2\Delta_l+2\Delta_{l'}-3}.
   \ee
   
  As mentioned above, for higher partial waves of $\psi$, contributions due to the exchange of other intermediates would also arise. The breaking of conformal invariance can be neglected for these contributions leading to the behaviour
  \be
  \label{behav}
  <O_{l_1}(\omega_1)O_{l_2}(\omega_2)O_{l_3}(\omega_3)^\dagger O_{l_4}(\omega_4)^\dagger> \sim e^2 \pqty{{r_h^2\over G_N}}\, \omega^{\sum_{i}^4 \Delta_i-2}
  \ee
By comparing, eq.(\ref{fm2}) and eq.(\ref{behav}) we see that $S$-wave exchange dominates, when it is allowed, at sufficiently low frequencies, meeting the condition,
\be
\label{condaa}
\omega \ll \frac{r_h}{L_2^2},
\ee  where we have inserted factors of $L_2$ as required by dimensional analysis. 

This is the same condition as eq.(\ref{condo2}).

In general,  several higher partial waves can contribute in the intermediate  channel. As a result,  there can be an intermediate range of frequencies    where the cumulative contribution of the higher partial wave exchange is significant. This happens  if about\footnote{The higher partial waves are further suppressed because they correspond to operators with higher anomalous dimensions, as we saw in section \ref{scalarsection}. We neglect this effect in the estimate below.} 
\be
\label{hpc}
n\sim \frac{r_h}{L_2^2 \omega}
\ee
number of partial waves become important. 

Let us examine if  eq.(\ref{hpc}) can be met   while keeping the mass of the KK mode intermediate states $m_{KK}$   to be no   bigger than of order 
unity  in units of $L_2$, and therefore their anomalous dimensions  to be of order unity at most.   The discussion in this section generalises in a straightforward way to the $AdS_2\times S^d$ case, see section \ref{nads}.  We might as well  analyse this more general case. 
The number of KK modes with $m_{KK}\le O(\frac{1}{L_2})$ is 
\be
\label{condd}
n_{KK}\sim \pqty{{r_h\over L_2}}^{d}.
\ee
 Thus eq.(\ref{hpc}) is met if
 \be
 \label{condfd}
 \omega r_h \pqty{{r_h \over L_2}}^{d-2} \sim O(1).
 \ee
 From eq.(\ref{condo2}) and eq.(\ref{condfd}) we learn that for $\omega$ lying in the range
 \be
 \label{condeea}
 {1\over r_h}\pqty{{L_2\over r_h}}^{d-2} \ll \omega \ll {r_h\over L_2^2},
 \ee
 the effect of the higher partial waves will be more significant than the $S$-wave sector while for  smaller $\omega$ the $S$ wave sector will dominate. 
 For $d=2$, we see that eq.(\ref{condeea}) can be met  for a big black hole with $\frac{r_h}{L_2}\gg 1$, while  for $d>2$ it becomes even easier to meet the condition for such black holes.  
  
  We will also discuss black holes in asymptotically flat space later in section \ref{nads}; their behaviour is similar to small black holes in $AdS$ space. The formulae above (expressed as given in terms of the parameters, $r_h, L_2$)  are applicable here as well, with $r_h=L_2$.  The condition eq.(\ref{condo2}) for the near $AdS_2$ analysis to apply is more appropriately stated as eq.\eqref{condff} for a probe of mass $m$ and eq.(\ref{condeea}) then takes the form 
  \be
  \label{condefl}
  {1\over r_h}\pqty{{L_2\over r_h}}^{d-2} \ll \omega \ll { m r_h\over L_2}.
  \ee
   Setting $L_2=r_h$ this becomes
   \be
   \label{condefg}
   {1\over r_h}\ll \omega \ll m.
   \ee
   This condition can also  be met for a sufficiently big black hole. 
\section{Charged Matter Coupled To Gravity and Background Gauge Field}
\label{bgf}
Here we consider the effects of gravity  and the coupled gauge field, which is turned on in the background extremal solution,  for  calculating the  four-point function of  a scalar field charged under this gauge field.  The system is given by the Euclidean version of the action, eq.\eqref{act1}, coupled to the  scalar,
\begin{align}
S_{matter}=\frac{1}{G_N}\int d^4x\,\sqrt{g}\,\pqty{|D\psi|^2+m^2|\psi|^2}\label{prob}.
\end{align}
We are interested in the behaviour at low frequencies,  eq.(\ref{condo2}.

We begin the analysis by first considering the $S$-wave sector. We will see that this  sector  dominates the behaviour  at sufficiently low frequencies.
This is due to both contributions from the time reparametrisation mode  and a phase mode analogous to the one described above. Correlation functions for a field uncharged under the background gauge field were already discussed in \cite{Nayak:2018qej}. 

\subsection{The Two Point Function}

Let us first examine the behaviour of the scalar field in the near-horizon region. 
To leading order, it satisfies the equation, 
\be
\label{eomcs}
\bar{D}^{\mu}\bar{D}_{\mu}\psi-m^2\psi=0,
\ee
where ${\bar D}_{\mu}=\del_{\mu}-i \,e\,\bar{A}_{\mu}$, with ${\bar A}_\mu$ being the background gauge field, $e$ being the charge of the probe scalar\footnote{In our conventions the gauge potential $A_\mu$ is dimensionless and $e$ has dimensions of $(length)^{-1}$.} and the indices in eq.(\ref{eomcs}) being raised by the background metric. 
We work in the gauge where only the ${\bar A}_t$ component of the background gauge field is non-zero. In the  $AdS_2$ region it is given by  
\begin{equation}
\label{abart}
\bar{A}_t=\frac{L_2^2}{r_h^2}\,\frac{Q}{z}.
\end{equation}
As a result, is it easy to see that the ``effective mass" of $\psi$  is 
\begin{equation}
m_{eff}^2=m^2+\frac{e^2Q^2L_2^2}{r_h^4}\label{meff}.
\end{equation}
Let us take the boundary of the $AdS_2$ region to be at $z=\delta$, where $z$ is defined in eq.\eqref{zdef}. Then from eq.(\ref{eomcs}), eq.(\ref{meff}), we see that the asymptotic behaviour as $z\rightarrow \delta$ is 
\be
\label{asbeh}
\psi(t,z)=z^{\tilde{\Delta}_-}\,\psi(t)\,+\,c\,(z^{\tilde{\Delta}_+}-z^{\tilde{\Delta}_-}\,\delta^{\tilde{\Delta}_+-\tilde{\Delta}_-})\int dt'\frac{\psi(t')}{|t-t'|^{2\tilde{\Delta}_+}},
\ee
with 
\be
\label{defdpm}
\tilde{\Delta}_{\pm}=\half\pm\sqrt{\quarter+m^2L_2^2+\frac{e^2Q^2L_2^4}{r_h^4}}.
\ee
$\psi(t)$ is the value for the scalar field at the $AdS_2$ boundary. 

It then follows that the two-point function for the scalar at low energies is given by 

 \be
\label{2pt}
S_{2pt}= -\pqty{\frac{r_h^2}{G_N}}\,4\pi \,c \, (\tilde{\Delta}_+ - \tilde{\Delta}_-) \int dt\, dt' \,\frac{\psi^\dagger (t) \psi(t')}{|t-t'|^{2\tilde\Delta_+}}.
\ee
We see that the scalar corresponds to an operator  of dimension $\tilde{\Delta}_+$. 

\subsection{The Four Point Function}
\label{4ptsub}
We turn to the four-point function next. As mentioned above we are considering  the $S$-wave sector here. To begin, we work directly in the $4$ dimensional theory. 

The scalar, once turned on, gives rise to a stress tensor and charged current which in turn perturb the metric and the gauge field, giving rise to a four-point function.
As discussed in appendix \ref{4dbgf}, the resulting on-shell action  to the required order for the four-point function is,
\begin{align}
S_{OS}&=-32\pi^2 G_N \int dt\, dr\, \left(\frac{2a^2b^3}{b'}\,T_{rr}\frac{1}{\del_t}T_{tr}^{\psi}-a^2b^2\Big(1+\frac{2a'b}{b'a}\Big)\,T_{tr}\frac{1}{\del_t^2}T_{tr}^{\psi}\right)\nonumber\\
&\qquad-\frac{2\pi\,e^2}{G_N}\int dt\,dr\,b^2\,a^4\,J_r\,\del_t^{-2}J_r-\frac{2\pi\,e}{G_N}\int dt\,dr\,b^2\,a^2\,\delta A_r\,J_r\label{ost}.
\end{align}
Here $T_{\alpha \beta}^{\psi}$ is the contribution to the stress tensor from the scalar $\psi$, eq.\eqref{mstress}, and $T_{\alpha\beta}$ has contributions from both the scalar as well as the gauge field perturbation, expanded to the required order for the four-point correlator,  see eq.\eqref{fstress}. The current $J_r$  is quadratic in $\psi$, see eq.\eqref{current}, and the gauge field perturbation $\delta A_r$ is given in terms of the metric perturbations in eq.\eqref{gaugepert}.

We have seen, both in the uncharged case, and the probe gauge field case above, that an $O(r_h^3)$ contribution is expected to arise in the four-point function.
For the terms in the first line of eq.(\ref{ost}) which are quadratic in the stress tensor, such  contributions do arise  and they get support from the near horizon region. The near horizon contribution can be expressed as an integral on the boundary of the near $AdS_2$, as discussed in eq.(\ref{term11bdy}) and takes the form, 
\begin{align}
S_{bdy}^{\psi}=-\frac{32\pi^2 G_N r_h^3}{L_2^2}\int dt\, ({\tilde T}^\psi_{tz}-z \partial_t {\tilde T}^\psi_{zz})\,\frac{1}{\del_t^4}\,(\tilde{T}_{tz}^\psi-z \partial_t \tilde{T}_{zz}^\psi).\label{Ttperm}
\end{align} 
Using the fact that  
\be
\label{deftildet}
{\tilde T}^\psi_{\alpha\beta}=\frac{1}{2G_N}\pqty{\del_{\alpha}\psi^\dagger\del_{\beta}\psi+\del_{\beta}\psi^\dagger\del_{\alpha}\psi-\bar{g}_{\alpha\beta}\,(\bar{g}^{\mu\nu}\del_{\mu}\psi^\dagger\del_{\nu}\psi+m^2\psi\psi^\dagger)},
\ee
and the asymptotic form for $\psi$, eq.(\ref{asbeh}), the expression eq.\eqref{Ttperm} can be expressed in terms of the value $\psi$ takes at the boundary of the near $AdS_2$ region to give, 
\begin{align}
S_{bdy}^{\psi}&=-32\pi^2\,c^2\,(\tilde{\Delta}_+-\tilde{\Delta}_-)^2\,\tilde{\Delta}_+^2\,\frac{r_h^2}{G_N}\,\frac{r_h}{L_2^2}\nonumber\\
&\int dt\,dt'\,dt''\,\Bigg(\quarter\, F_1(t,t')\frac{1}{\del_t^4}F_1(t,t'')-\tilde{\Delta}_-\,F_1(t,t')\frac{1}{\del_t^4}F_2(t,t'')+\tilde{\Delta}_-^2\, F_2(t,t')\frac{1}{\del_t^4}F_2(t,t'')\Bigg)\label{sbdypsi},\\
F_1(t,t')&=\frac{\del_t\psi^\dagger(t)\psi(t')+\del_t\psi(t)\,\psi^{\dagger}(t')}{|t-t'|^{2\tilde{\Delta}_+}},\nonumber\\ F_2(t,t')&= \frac{1}{2\tilde\Delta_+}\pqty{\psi^\dagger(t)\psi(t')+\psi(t)\,\psi^{\dagger}(t')}\,\del_t|t-t'|^{-2\tilde{\Delta}_+}\label{f12}.
\end{align}

From the  first term in the second line of eq.(\ref{ost}), which is quadratic in the current $J_r$, we again get a contribution of $O(r_h^3)$. This can be analysed in a manner completely analogous to the probe gauge field case in section \ref{probeu1}. The resulting answer takes the form in eq.\eqref{final4p} with $\Delta _\pm$ replaced by ${\tilde \Delta}_\pm$ and is given by

\begin{align}
  S_{J}&= \pqty{\frac{r_h^2}{G_N}} \,r_h\,2\pi\,c^2\,e^2\, (\tilde\Delta_+-\tilde\Delta_-)^2\nonumber\\
  &\quad\qquad\times\int dt\, dt' dt''\pqty{\frac{\psi(t)\psi^{\dagger}(t'')-\psi(t'')\psi^{\dagger}(t)}{|t-t''|^{2\tilde\Delta_+}}}\frac{1}{\del_t^2}\pqty{\frac{\psi(t)\psi^{\dagger}(t')-\psi(t')\psi^{\dagger}(t)}{|t-t'|^{2\tilde\Delta_+}}}.\label{sj}
\end{align}
 
Finally, it can be shown that the last term in eq.\eqref{ost} involving the gauge field perturbation $\delta A_r$  cancels with a corresponding term in the first line of eq.\eqref{ost}, see appendix \ref{4dbgf} for details. Thus $S_{bdy}^{\psi}+ S_J$ is the total answer for the four-point function in the $S$-wave sector.

\subsection{The Action For Time Reparametrisation and Phase Modes} 
\label{modeactsub}

It is useful at this stage to consider the two dimensional theory obtained by $S$-wave reduction of the  $4$ dimensional theory, eqs.\eqref{act1} and \eqref{prob}.

Expanding the dilaton field $\Phi$ (defined in eq.\eqref{dimmet} of appendix) to linear order,
\be
\label{linordil}
\Phi=r_h(1+\phi)
\ee
we get, from an analysis similar to section 5 of \cite{Nayak:2018qej}  that 
 in the near-horizon region the two dimensional theory has an action\footnote{We work in a gauge where the perturbation $\delta A_\alpha$ satisfies the boundary condition $F^{\alpha\beta}\delta A_\alpha n_\beta=0$, with $n_\beta$ being the normal to the boundary.}
 (see appendix \ref{2dbgf})
 \begin{align}
 S&=-\frac{r_h^2}{G_N}\pqty{\quarter\int d^2x\,\sqrt{g}\,R+\half \int_{\del}\sqrt{\g}\,K}-\frac{r_h^2}{G_N}\,\frac{2\pi}{r_h}\int dt\,\dot{\bar\theta}^2\nonumber\\
 &\qquad-\frac{r_h^2}{G_N}\,\pi\int d^2x\,\sqrt{g}\,(F^2-F_Q^2)-\frac{r_h^2}{G_N}\,3\pi\int d^2x\,\sqrt{g}\,\phi\,(F^2-F_Q^2)\nonumber\\
 &\qquad-\frac{r_h^2}{G_N}\,\half\int d^2x\,\sqrt{g}\,\phi\, (R-\Lambda_2)-\frac{r_h^2}{G_N}\int_{\del}\sqrt{\gamma}\,\phi\,K+\frac{r_h^2}{G_N}\,\frac{1}{L_2}\int_{\del}\sqrt{\g}\,\phi\label{final2d}.
 \end{align}

Here we have added boundary terms and an extra boundary  counterterm, \cite{Nayak:2018qej}. $F^Q_{\alpha\beta}$  above is given by 
\begin{align}
F_{rt}^Q=\frac{Q}{r_h^2\,\sqrt{g}\,g^{rr}g^{tt}}\label{fqrt},
\end{align}
and $\dot{\bar \theta} $ is given  by eq.\eqref{defthetabar}. 

We see that the resulting two dimensional theory is closely related to the JT theory of gravity in the presence of an extra gauge field. In the absence of any sources the theory admits a solution with $\phi=0$, and with the metric being $AdS_2$, and the gauge field taking the form, 
\begin{equation}
\label{gaugesol}
F_{rt}=F^Q_{rt}.
\end{equation}
It is easy to verify that the  resulting  stress tensor of the gauge field cancels the stress tensor which arises from the $F_Q$ term. 

We can now examine the dynamics of the near-horizon region in this two dimensional theory. The dynamics includes fluctuations of the boundary of this region  corresponding to time reparametrisations\footnote{There are also additional effects since the geometry is no longer $AdS_2$ once  $F$ departs from the  value $F^Q$.}
and on general arguments in \cite{Maldacena:2016upp,Nayak:2018qej} leads to the Schwarzian action,
\begin{align}
S_{sch}=-\frac{r_h^2}{G_N}\,\frac{L_2^2}{r_h}\int_{bdy} Sch\pqty{f(t)},\label{finsch}
\end{align}
where
 \begin{align}
 Sch\pqty{f(t)}  &=   -{1\over 2 } {(f'')^2\over (f')^2}+ \pqty{{f''\over f'}}'\label{schdef}.
 \end{align}
 Writing
 \be
 \label{deff}
 f(t)  =  t+ \epsilon(t),
 \ee
 and expanding to quadratic order in $\epsilon$, eq.\eqref{finsch} becomes

 \be
\label{schwa}
S_{\epsilon}=-\frac{r_h^2}{G_N}\,\frac{L_2^2}{r_h}\int dt \pqty{-{3\over 2 }(\epsilon'')^2-\epsilon'\epsilon'''}.
\ee
As noted in \cite{Maldacena:2016upp,Nayak:2018qej}, the coefficient in eq.(\ref{schwa}) leads  to the correct   near-extremal free energy  in the Canonical Ensemble. 

Similarly, a  phase mode corresponding to the gauge field perturbation
\begin{equation}
\label{gpert}
\delta A_r=\frac{1}{\sqrt{g_4}\,g^{rr}g^{tt}}\,\delta\theta(t),
\end{equation}
must arise to account for the Coulomb ``dressing" of perturbations. Analogous to the phase mode in section \ref{probeu1} 
we get an action for this mode to be,
\begin{equation}
\label{sphase}
S_{phase}=-\frac{r_h^2}{G_N}\,2\pi\,\frac{1}{r_h}\int dt\,\delta\dot{\theta}(t)^2.
\end{equation}
The coefficient  in eq.(\ref{sphase}) is related, upto a factor of $2$, to the coefficient of the term required to connect the canonical and grand canonical partition functions \cite{Hawking:1995ap} as discussed in appendix \ref{2dbgf}, eq.\eqref{actpb}. Note that both terms in eq.\eqref{schwa} and eq.\eqref{sphase} are suppressed by a factor of $\frac{1}{r_h}$. 

The coupling between $\epsilon$ and $\delta \theta$ can be obtained from symmetry considerations, \cite{Davison:2016ngz} as explained in appendix \ref{2dbgf}.
This leads to a full action for $\epsilon $ and $\delta \theta$, which upto quadratic order in $\epsilon$ is given by 
\begin{equation}
\label{modeact}
S_{mode}=-\frac{r_h^2}{G_N}\,2\pi\,\frac{1}{r_h}\int dt\,\pqty{\delta\dot{\theta}(t)+\dot{\epsilon}(t) \dot{\bar{\theta}}}^2+\frac{r_h^2}{G_N}\,\frac{L_2^2}{2}\,\frac{1}{r_h}\int dt\,\epsilon(t)\,\epsilon^{(4)}(t).
\end{equation}

We can now consider coupling these degrees of freedom to the source terms which arise from the scalar $\psi$. 
Carrying out a time reparametrisation on  the two-point function, eq.\eqref{2pt},  gives rise to a coupling to $\epsilon(t)$, 
\begin{align}
S_{coup1}=\frac{r_h^2}{G_N}\,4\pi c\,(\tilde{\Delta}_--\tilde{\Delta}_+)\,\tilde{\Delta}_+\int dt\,dt'\,\pqty{\frac{\psi^\dagger(t)\psi(t')+\psi^\dagger(t')\psi(t)}{|t-t'|^{2\tilde{\Delta}_+}}}\pqty{\dot{\epsilon}(t)-\frac{2\epsilon(t)}{|t-t'|}},\label{scoup1}
\end{align}
Similarly a phase rotation and further time reparametrisation gives rise to the coupling 
\begin{align}
S_{coup2}=\frac{r_h^2}{G_N}\,4\pi c\,e\,(\tilde{\Delta}_+-\tilde{\Delta}_-)\,i\int dt\,dt'\,\pqty{\frac{\psi(t)\psi^{\dagger}(t')-\psi(t')\psi^{\dagger}(t)}{|t-t'|^{2\tilde{\Delta}_+}}}(\delta\theta(t)+\epsilon(t)\dot{\bar{\theta}})\label{scoup2}.
\end{align}

The response to the sources can now be calculated and gives rise to a four-point function which is exactly the sum of the two terms, eq.\eqref{sbdypsi} and \eqref{sj} as discussed in appendix \ref{2dbgf}.

We have focused on the zero temperature case above. At finite temperature the action for the time reparametrisation modes is given by
\begin{align}
 S_{sch}&=-\frac{r_h^2}{G_N}\,\frac{L_2^2}{r_h}\int_0^\beta d\tau\, Sch\pqty{\frac{\beta}{\pi}\,\tan{\pi \over \beta} (\tau + \epsilon)}\nonumber\\
 &= -\frac{r_h^2}{G_N}\,\frac{L_2^2}{r_h}\int_0^\beta d\tau \pqty{-{3\over 2 }(\epsilon'')^2-\epsilon'\epsilon'''+\frac{2\pi^2}{\beta^2}(\epsilon')^2}\label{actschwa}
 \end{align}
 where $\tau \simeq \tau+\beta$ and $\epsilon'={d\epsilon \over d\tau}$. In going to the second line, we have expanded the action to quadratic order in $\epsilon$.
 And for the phase mode, eq.\eqref{sphase} takes the form, 
 \begin{align}
 S_{phase}=-\frac{r_h^2}{G_N}\,2\pi\,\frac{1}{r_h}\int_0^\beta d\tau\,\pqty{\frac{d\delta\theta}{d\tau}+\dot{\bar{\theta}}\,\frac{d\epsilon}{d\tau} }^2,\label{phasetemp}
 \end{align}
 where ${\dot {\bar \theta}}$ continues to be given by eq.\eqref{defthetabar}. The coupling between the time reparametrisation and phase modes to the field $\psi$ at finite temperature can be obtained by doing a phase rotation and time reparametrisation $t=\frac{\beta}{\pi}\,\text{tan}\frac{\pi}{\beta}(\tau+\epsilon)$ to the two point function. This gives the couplings for the time reparametrisation mode $\epsilon(\tau)$ and the phase mode $\delta \theta(\tau)$ with matter to be
 \begin{align}
 S_{coup1}&=\frac{r_h^2}{G_N}\,4\pi c\,(\tilde{\Delta}_--\tilde{\Delta}_+)\,\tilde{\Delta}_+\,\pqty{\frac{\pi}{\beta}}^{2\tilde{\Delta}_+}\int d\tau_1\,d\tau_2\,\pqty{\frac{\psi^\dagger(\tau_1)\psi(\tau_2)+\psi^\dagger(\tau_2)\psi(\tau_1)}{\text{sin}^{2\tilde{\Delta}_+}\frac{\pi}{\beta}(\tau_1-\tau_2)}}\nonumber\\
 &\qquad\qquad\qquad\qquad\qquad\times \pqty{\epsilon'(\tau_1)-2\epsilon(\tau_1)\,\text{cot}\frac{\pi}{\beta}(\tau_1-\tau_2)},\label{scoup1t}\\
 S_{coup2}&=\frac{r_h^2}{G_N}\,4\pi c\,e\,(\tilde{\Delta}_+-\tilde{\Delta}_-)\,\pqty{\frac{\pi}{\beta}}^{2\tilde{\Delta}_+}\,i\int d\tau_1\,d\tau_2\,\pqty{\frac{\psi(\tau_1)\psi^{\dagger}(\tau_2)-\psi(\tau_2)\psi^{\dagger}(\tau_1)}{\text{sin}^{2\tilde{\Delta}_+}\frac{\pi}{\beta}(\tau_1-\tau_2)}}\nonumber\\
 &\qquad\qquad\qquad\qquad\qquad\qquad\qquad\qquad\qquad\times\pqty{\frac{d\delta\theta}{d\tau_1}+\dot{\bar{\theta}}\,\frac{d\epsilon}{d\tau_1} }\label{scoup2t}.
 \end{align}

 The terms eqs.\eqref{actschwa}, \eqref{phasetemp}, \eqref{scoup1t} and \eqref{scoup2t} then give rise to the four-point function at finite temperature, and also the higher point functions which can be obtained through tree level exchange of these modes.

\subsection{Additional Comments}
We have been considering the scalar field to be in the $S$-wave sector so far, and have analysed the effects of gravity and the gauge field which can  be taken to be  in the $S$-wave sector as well. Let us now consider higher partial waves for the scalar field. 
To analyse the resulting behaviour it is useful to carry out a KK reduction on the $S^2$. From gravity we then get $3$ additional massless gauge fields in  the $2$ dimensional theory which arise  from the $S^2$ isometries. In addition, we get  extra  massive fields, which can be of spin $0,1,2,$ arising from the gravity and gauge perturbations which are coupled.

The contribution to the four-point function from the exchange of the massive degrees can be obtained by working in the $AdS_2$ geometry, and the breaking of the $AdS_2$ isometry due to the back reaction of the dilaton can be neglected in this sector. These contributions go like $O(r_h^2)$ and are analogous to the contributions studied in section \ref{scalarsection} due to the exchange of an intermediate scalar. 

The contribution due to the additional massless gauge fields will be analogous to the those studied in section \ref{probeu1} due to a probe gauge field. The higher partial waves of the scalar will carry charge under these gauge fields. The resulting contribution will  be of $O(r_h^3)$ and of the form, eq.(\ref{fm2}), with the anomalous dimensions $\Delta_i$ and charges
being determined by the angular momentum of the scalar partial waves. 

At low energies the $O(r_h^3)$ contributions, if present,  will dominate in the four-point function. These will arise from both the $S$-wave sector and the additional KK massless gauge fields. The additional massive field discussed above will each contribute of $O(r_h^2)$, and will be suppressed at low energies. However, their cumulative contribution could be significant when the energy is not very small, see subsection \ref{adcom} for related discussion. 

We have focussed on the $4$ point function above, higher point correlators can be obtained through successive tree level exchange of the time reparametrisation and phase modes.

 \section{The Near-$\mathbf{AdS_2}$ Near-$\mathbf{CFT_1}$ Correspondence}
 \label{nads}
 The discussion above leads us to a formulation of the near $AdS_2$ near $CFT_1$ correspondence (henceforth referred to as the $NAdS_2/NCFT_1$ correspondence).
 We have been analysing a four dimensional theory above. However,  we could start with a general higher dimensional situation as well \footnote{ Asymptotic  $AdS_3$ space 
 can be  different, we therefore restrict ourselves to four and higher dimensions.}. The near-horizon geometries to which our discussion applies  have  the form\footnote{We have not considered rotating extremal black holes in this paper.} $AdS_2\times S^d$. 
 The resulting low energy dynamics can be analysed by constructing the two dimensional theory obtained by KK reduction on $S^d$. This will contain two dimensional gravity, coupled to the dilaton, which is the radius of the $S^d$,  and the gauge fields turned on in the near-horizon region. In addition there will be fields which are not excited in the background. In general this will include massless gauge fields and additional massive degrees of freedom \footnote{Massless scalars need to be analysed separately.}.

 For concreteness let us consider the case of a single background gauge field in asymptotically $AdS_{d+2}$ space.  We have seen  that the $S$-wave sector determines the low energy response of the system. In particular, working in the Canonical Ensemble at fixed charge,  the Schwarzian action, eq.\eqref{schwa},  gives rise to  the  free energy  at small temperatures, $T\ll \mu$. 
 In addition this  action along with  the phase mode, eq.\eqref{modeact}, gives rise to the four and higher point correlators for probes at low energies, $\omega \ll \mu$. 
 For the additional massless gauge fields in the two dimensional theory (including those arising from the isometries of $S^d$) there will be additional phase modes whose effects also need to be incorporated, as discussed above. The action for the time reparametrisation and phase modes will be suppressed by a power of $\frac{1}{r_h}$, where $r_h$ is the scale characterising the departures from the $AdS_2$ geometry which  is related to the chemical potential for the gauge field by $r_h\sim \mu L_{d+2}^2$. As a result, the four-point function for example, of a probe will be enhanced by an additional power of ${ r_h \over \omega  L_{d+2}^2}$. The effects of massive fields, of various spins, can be incorporated in a straightforward manner as discussed above, and the departure from the $AdS_2$ geometry can be neglected in calculating these effects at low energies. The contributions due to the individual exchange of each massive mode will be suppressed, but cumulatively could be significant, see subsection \ref{adcom}. 
 
 One comment is worth making before we proceed. The same length scale $r_h$ enters in the discussion of the infrared dynamics in two conceptually different ways in the example above of an extremal RN black hole in $AdS_{d+2}$. First, $r_h$ is the radius of the horizon and therefore determines the ground state entropy. Second, it also characterises the scale at which corrections to the $AdS_2\times S^d$ geometry become significant. It is helpful to keep this  distinction in mind while analysing the low energy dynamics.

 The essential points in the discussion above for the asymptotically $AdS_{d+2}$ case are in fact valid more generally in any situation   where the near-horizon geometry takes the form $AdS_2\times S^d$. This includes   extremal black holes in asymptotically flat space-time whose behaviour is similar to ``small" black holes in $AdS_{d+2}$ space. 
 
 It is worth commenting on this case in some more detail here.  For the asymptotically flat case,  the same scale $r_h$  in fact plays three different roles. It is the  radius of the near horizon $AdS_2$, $L_2$,  as well as    the radius of the $S^d$. In addition,  it also characterises the  scale at  which corrections to the $AdS_2$ geometry become significant. 
 Once one keeps track of this distinction   carefully, all formulae for the asymptotically $AdS_{d+2}$ case generalise and can be applied to this case as well. 
 More explicitly, to apply  the formulae obtained in  the discussion of the asymptotically $AdS_{d+2}$ case  for the asymptotically flat case, one should  express them first  in terms of $r_h$ (which enters in the $AdS_{d+2}$ case both as the  horizon value of the $S^d$, and the scale characterising the departure from the IR conformal symmetry) and $L_2$, and thereafter  set $L_2=r_h$.
 
 In fact, for the case of four dimensions, we have already been careful to present the formulae for the $AdS_{d+2}$ case keeping these aspects in mind. The Schwarzian term and action for the phase modes as given by eq.\eqref{modeact} are  also valid for the asymptotically flat case, with $L_2=r_h$. Coupling them to the matter fields, as given in eqs.\eqref{scoup1} and \eqref{scoup2} then gives the correlation functions which characterise the response of the black hole in the four dimensional asymptotically flat case as well.

  It is worth pointing out that the chemical potential of the extremal black hole in the asymptotically flat case   is  different from the $AdS_{d+2}$ case  and is of order unity (in our units) instead of being of  order $\mu \sim \frac{r_h}{L_2^2}$ as noted in eq.\eqref{chempot}. 
    In addition,  since $L_2$ is determined by the charge of the black hole in the asymptotically flat case it is more appropriate to think of the mass of the probe field as being an independent parameter. The condition for being at small enough energies, so that  the resulting dynamics is described by the near-$AdS_2$ region, is then more appropriately stated as eq.\eqref{condff}, which becomes, 
  \be
  \label{condeae}
  \omega \ll m.
  \ee
  The requirement of being at low temperatures is  given by $T\ll \frac{r_h}{L_2^2}=\frac{1}{r_h}$ and the free energy continues to be  given by 
  eq.\eqref{free} and is correctly obtained from the Schwarzian term, eq.\eqref{schwa}.

 In this way  one can  apply the discussion developed above equally well to asymptotically flat case. In summary,  the low energy behaviour arises from the near-$AdS_2$ dynamics, and is dominated at sufficiently low energies by the time reparametrisation and phase modes whose action is suppressed by the scale $r_h$ which governs the corrections to the near-horizon geometry. 
 
 One expects that the near-$AdS_2$ dynamics which can be calculated as given above should correspond to the low energy behaviour of an approximate one dimensional CFT, in any consistent theory of gravity.  Different black holes should  correspond to different  CFTs but  regardless of microscopic difference their dynamics, at low energies, should be the same and determined by the time reparametrisation and phase modes whose action arises due to effects which break the conformal symmetry.

 One can further extend these considerations to a general black hole carrying multiple charges. It is well known that extremal black holes in general exhibit the attractor mechanism, which is a general feature of supersymmetric and non-supersymmetric extremal black holes \cite{Ferrara:1995ih,Strominger:1996kf,Ferrara:1996dd,Ferrara:1996um,LopesCardoso:1998tkj,LopesCardoso:1999cv,LopesCardoso:1999fsj,LopesCardoso:2000qm,Ooguri:2004zv,Sen:2005wa,Goldstein:2005hq,Kallosh:2005ax,Astefanesei:2006sy,Dabholkar:2006tb,Ferrara:1997tw,Kallosh:2006bt,Alishahiha:2006ke}. Moduli fields, which are generically present in string theory, are fixed at the horizon by the charges, and the entropy as a result is only a function of the charges and not the asymptotic values the moduli take. One the other hand the deviation away from the $AdS_2$ region depends on the asymptotic value for these moduli, which determine how the attractor values are approached at the horizon. As a result,  the ground state entropy is  universal and determined by the charges alone, while we expect  the Schwarzian and phase actions to have coefficients which are dependent  on additional data pertaining to the asymptotic behaviour of the moduli which determine how the IR theory  is approached from the UV. A more detailed analysis of the connections between the attractor phenomenon and the low energy dynamics of extremal black holes is left for the future.

 \section{Conclusions}
 We have analysed the behaviour of charged extremal and near-extremal black holes in this paper, continuing the analysis begun in \cite{Nayak:2018qej}. 
 In particular, we have considered charged matter fields and studied their correlation functions showing that at low energies the leading contribution arises  due to the effects of the time reparametrisation mode and an additional phase mode. These effects are enhanced at low frequencies by a factor of ${r_h\over L_2^2 \,\omega}$ where $r_h$ is the scale characterising departures from the $AdS_2$ geometry and $L_2$ is the  radius of the $AdS_2$ geometry. The action governing these modes at finite  temperatures is given by
 eqs.\eqref{actschwa} and \eqref{phasetemp}.
 It is known that the Schwarzian term correctly captures the near-extremal specific heat in the canonical ensemble. We have also shown that the action eqs.\eqref{actschwa} and \eqref{phasetemp} gives rise to the four-point function for the charged scalar field.
 
 While our discussion has focussed on the four-point function, it can be easily extended to higher point correlators which can be calculated via succesive tree level exchanges of the time reparametrisation and phase modes. 
 Also, some of the formulae  in the discussion on correlators above were  written down for  the extremal case, however they can easily be extended to the case of non-extremal black holes at low temperatures as discussed at the end of subsection \ref{modeactsub}.

 It is also worth mentioning that the phase mode we find does not correspond to a gauge transformation in the $AdS_2$ space-time. For example, from eq.(\ref{formar}) we see that the field strength is non-zero since $\theta$ is time dependent. This is in contrast to the time reparametrisation mode which corresponds to asymptotic isometries in the $AdS_2$ space-time.  Physically speaking, this difference arises because a spherically symmetric configuration  of charged matter gives rise to an electric field, while a finite temperature black hole in $AdS_2$ is still described locally by the $AdS_2$ geometry. 
  
 We have also studied  the behaviour beyond the $S$-wave sector and found that it can be analysed by carrying out a KK reduction down to the near $AdS_2$ theory. Additional massless gauge fields which arise from isometries in the KK reduction give rise to additional phase modes which give enhanced effects. Massive modes arising  in the KK reduction give rise to effects which are not enhanced and which can be calculated in the $AdS_2$ geometry neglecting any corrections to it. While the effects of each massive mode is small, they can cumulatively have a significant effect.

 We have also discussed how the analysis carried out for big black holes in  asymptotically $AdS$ space can be easily applied to more general situations, including asymptotically flat black holes which give rise to a near-horizon geometry of the $AdS_2\times S^d$ form. 
 In  the  general case of multi-charge black holes, the attractor mechanism leads to the ground state entropy being universal, however,   the departure from the $AdS_2$ region  depends  on the asymptotic values moduli take and this could affect the coefficient of the Schwarzian  and the phase mode action. 
 
 In this way, our analysis allows us to formulate a general method  for calculating the low energy behaviour of extremal black holes. This should correspond on general grounds to the low energy behaviour of microscopic dual theories which are approximate one-dimensional CFTs,  with  the breaking of conformal invariance being important in giving rise to the action for the time reparametrisation and phase modes.
 
 We have not considered rotating Kerr black holes here. These are charged with respect to the KK gauge fields which arise from the isometries of  $S^d$.
 We hope to return to this question in a subsequent paper, see \cite{Castro:2018ffi} for some recent discussion.  
 
 There are many interesting microscopic models of extremal black holes in string theory, arising for example from $D$-brane constructions. It will be worthwhile applying the results of this paper to these models to see  if more can be learnt about their dynamics. 
 
 Our results also provide good motivation for the study of the Jackiw--Teitelboim model of gravity, coupled to additional fields, in more detail, since we see that this model captures the low energy behaviour of extremal systems in many cases. We hope to report on some progress in this direction as well in the future.


\acknowledgments
We thank   Gautam Mandal, Shiraz Minwalla, Subir Sachdev, Ashoke Sen, Ronak Soni and Spenta Wadia for insightful discussions and comments.
We are especially grateful to Juan Maldacena for his generous insights and comments. 
We acknowledge the support by the International Centre for Theoretical Sciences (ICTS) during a visit for participating in the program  ``AdS/CFT at 20 and Beyond" where part of this work was done. We also thank the participants of the program for their questions and comments. UM and VV thank the ``Infosys Foundation International Exchange Program of ICTS" for support in participating in Strings 2018 held in Okinawa, Japan.
UM gratefully acknowledges support from the Simons Center for Geometry and Physics, Stony Brook University for the 2018 Simons Summer Workshop, while this work was in progress.
We thank the DAE, Government of India, for support.  
SPT also acknowledges support from the J. C. Bose fellowship of the DST, Government of India. We also acknowledge support from the Infosys Endowment for Research on  the Quantum Structure of Spacetime. 
Most of all, we thank the people of India for generously supporting research in String Theory.

\appendix

\section{2-D Action for Charged Matter Coupled to a Probe U(1) Field}
\label{probe2du1}
In section \ref{probeu1} we considered a scalar field charged under a probe $U(1)$ field in the background of a RN black hole in $AdS_4$.
Here we give some details of the calculation of the onshell action eq.\eqref{final4p}.

We start by splitting the bulk integral in the 4D action eq.\eqref{actgf} into the near and far horizon regions,
\begin{align}
S_{gf}=-{1\over G_N}\int_{NH} d^4x \sqrt{g}\,\pqty{{1\over 4} F^2 + e\,A_\mu J^\mu}\,-\,{1\over G_N}\int_{FH} d^4x \sqrt{g}\,\pqty{{1\over 4} F^2 + e\,A_\mu J^\mu}.\label{gfactsplit}
\end{align}
The near horizon integral spans from the horizon to the boundary in the asymptotic $AdS_2$ region and the far horizon region lies between the $AdS_2$ boundary and the $AdS_4$ boundary.

Now consider the far horizon bulk integral in eq.\eqref{gfactsplit}. We can use the equations of motion eq.\eqref{maxres} to simplify it to obtain the onshell expression
\begin{align}
S_{FH}=-\frac{e}{2G_N}\int_{FH}d^4 x\,\sqrt{g}\,A_{\mu}J^{\mu}\,+\,\frac{1}{2G_N}\int_{\del AdS_2}d^3x\,\sqrt{\gamma}\,F^{\mu\nu}n_{\mu}A_{\nu},\label{osfh}
\end{align}
where the boundary term is evaluated at the $AdS_2$ boundary. Here $n_{\mu}$ is the normal vector to the $AdS_2$ boundary. Note that we ignore the AdS$_4$ boundary term since it does not contribute to the on-shell action. Also we work in the gauge where the only nonzero component is $A_r$.  We can take the
boundary to be located at a constant value of $r$, to the required order,   and hence we can ignore the  boundary term in eq.\eqref{osfh} as well.

Evaluating the bulk action in the far horizon region eq.\eqref{osfh} using the solution, eqs.\eqref{solg} and \eqref{solar}, we get
\begin{align}
S_{FH}&=-\frac{e}{2G_N}\int_{FH}d^4 x\,\sqrt{g}\,A_{r}J^{r}=-\frac{e^2}{2G_N}\int_{FH} d^4 x\,\sqrt{g}\,g_{tt}\,g^{rr}\,J_r\,\del_t^{-2}J_r\label{faronsh}.
\end{align}

We can obtain the solution for $\psi$ in the far horizon region from its behaviour at the $AdS_2$ boundary. In terms of $z$ coordinate eq.\eqref{zdef}, at the AdS$_2$ boundary $z=\delta$ we have,
\begin{equation}
\label{asymads2}
\psi(t,z)\big|_{\del AdS_2}=z^{\Delta_-}\,\psi(t)\,+\,c\,(z^{\Delta_+}-z^{\Delta_-}\,\delta^{\Delta_+-\Delta_-})\int dt'\frac{\psi(t')}{|t-t'|^{2\Delta_+}},
\end{equation}
where $c$ is a constant,
\begin{align}
c=2^{-2\nu}\,\frac{\Gamma(-\nu)}{\Gamma({\nu})},\quad \nu=\Delta_+-\Delta_-\label{cvalue}
\end{align}
and $\Delta_{\pm}$ are given in eq.\eqref{deltapm}.
In terms of $r$ coordinate, eq.\eqref{zdef}, the expression eq.\eqref{asymads2} becomes
\begin{equation}
\label{asymads2r}
\psi(t,r)\big|_{\del AdS_2}=\pqty{\frac{L_2^2}{r-r_h}}^{\Delta_-}\,\psi(t)\,+\,c\,\pqty{\pqty{\frac{L_2^2}{r-r_h}}^{\Delta_+}-\pqty{\frac{L_2^2}{r-r_h}}^{\Delta_-}\,\delta^{\Delta_+-\Delta_-}}\,\int dt'\frac{\psi(t')}{|t-t'|^{2\Delta_+}}.
\end{equation}
We can extend this solution beyond the $AdS_2$ boundary as follows. We are working with small values of $\omega$ which satisfy eq.\eqref{asb2}. Therefore, the $\omega$ dependence or equivalently the time dependence of the solution for $\psi$ is fixed to be that in eq.\eqref{asymads2r} for the region beyond the $AdS_2$ boundary where the condition eq.\eqref{asb2} is met. Therefore in this region, $\psi$ takes the form,
\begin{equation}
\label{fhsol}
\psi(t,r)=f_-(r)\,\psi(t)\,+\,c\,f_+(r)\,\int dt'\frac{\psi(t')}{|t-t'|^{2\Delta_+}},
\end{equation}
where the functions $f_{\pm}(r)$ are solutions to the equation
\begin{equation}
\label{fpmdiff}
\frac{1}{\sqrt{g}}\del_r\pqty{\sqrt{g}\,g^{rr}\del_rf_{\pm}}-m^2f_{\pm}=0.
\end{equation}
satisfying the condition that as $r\rightarrow r_c$ 

\begin{align}
f_-\big|_{\del AdS_2}& \rightarrow \pqty{\frac{L_2^2}{r-r_h}}^{\Delta_-},\nonumber\\
f_+\big|_{\del AdS_2}&\rightarrow \pqty{\pqty{\frac{L_2^2}{r-r_h}}^{\Delta_+}-\pqty{\frac{L_2^2}{r-r_h}}^{\Delta_-}\,\delta^{\Delta_+-\Delta_-}}\label{fbdy}.
\end{align}

Plugging eq.\eqref{fhsol} into the expression for the current eq.\eqref{current}, we get,
\begin{align}
\label{jsol}
J_r=-i\,c\,\int dt'\pqty{\frac{\psi(t)\psi^{\dagger}(t')-\psi(t')\psi^{\dagger}(t)}{|t-t'|^{2\Delta_+}}}\pqty{f_+\del_rf_--f_-\del_rf_+}.
\end{align}
The Wronskian in  eq.\eqref{fpmdiff},  can be shown to be of the form 
\begin{equation}
\label{wsol}
W[f_+,f_-]=f_+\del_rf_--f_-\del_rf_+=\frac{\tilde{c}}{\sqrt{g}g^{rr}}.
\end{equation}
Here $\tilde{c}$ is independent of $r$ and can be obtained by examining the behaviour of $W[f_+,f_-]$ near the $AdS_2$ boundary. Using the boundary condition eq.\eqref{fbdy} and the $AdS_2$ metric eq.\eqref{nearhor},
\begin{equation}
\label{ctild}
\tilde{c}=W\sqrt{g}\,g^{rr}=r_h^2 (\Delta_+-\Delta_-).
\end{equation}

Plugging in eqs.\eqref{jsol} and \eqref{wsol} into the action eq.\eqref{faronsh} and simplifying, we obtain
\begin{align}
\label{sfh}
S_{FH}&=\frac{r_h^4}{2G_N}\,(\Delta_+-\Delta_-)^2\,c^2e^2\int_{FH}d^4 x \,\frac{1}{\sqrt{g}}\,g_{tt}\, g_{rr}\,F(t)\del_t^{-2}F(t),
\end{align}
where $F(t)$ is defined as
\begin{equation}
\label{Fdef}
F(t)=\int dt'\pqty{\frac{\psi(t)\psi^{\dagger}(t')-\psi(t')\psi^{\dagger}(t)}{|t-t'|^{2\Delta_+}}}.
\end{equation}
Note that the metric is of the form
\begin{equation}
\label{metform}
ds^2=a(r)^2dt^2+a(r)^{-2}dr^2+r^2\,d\Omega_2^2,
\end{equation}
where $a^2$ is given in eq.\eqref{RNAdS4II}.
Therefore the integral over $r$ in eq.\eqref{sfh} becomes,
\begin{equation}
\label{intr}
\int_{r_c}^{\infty}\frac{dr}{r^2}=\frac{1}{r_c}.
\end{equation}
For small $\omega$ satisfying eq.\eqref{asb2}, we can approximate $r_c$ with $r_h$. This is because the integral over $r$ in the far horizon region spans till the location $r_c$ such that 
\begin{equation}
\frac{r_c-r_h}{L_2}\sim\omega L_2\Rightarrow r_c=r_h+O(\omega).\label{rccond}
\end{equation}
Therefore to leading order in $\omega$ after integrating over $r$ and also over $S^2$, the action in the far horizon region becomes
\begin{align}
\label{FHact}
S_{FH}=\frac{r_h^2}{G_N}\,2\pi\,r_h\,c^2\,e^2\,(\Delta_+-\Delta_-)^2\int dt \,F(t)\,\del_t^{-2}F(t),
\end{align}
where $F(t)$ is defined in eq.\eqref{Fdef}. We see that the region away from the horizon, near the $AdS_2$ boundary gives a contribution to the four-point function which is enhanced by a factor of $r_h$.
It is straightforward to see that the near horizon region does not give such an enhanced contribution to the four-point function; it contributes at $O(r_h^2)$.

\section{4-D Action Computation of $4$ point function in Background Gauge Field}
\label{4dbgf}
In this appendix, we give details of the $4$-point function  computation   of a scalar field charged under the gauge field which is excited in the background black hole solution. We work directly in $4$ dimensions in the $S$-wave sector. 

We start with the 4-D Euclidean action,
\begin{align}
S&=-\frac{1}{16\pi G_N}\int d^4x\,\sqrt{g}\,\pqty{R-2\Lambda}-\frac{1}{4 G_N}\int d^4x\,\sqrt{g}\,F^2\nonumber\\
&\quad+\frac{1}{G_N}\int d^4x\,\sqrt{g}\,\pqty{(D_{\mu} \psi)^{\dagger}D^{\mu}\psi+m^2\psi^{\dagger}\psi}-\frac{1}{8\pi G_N}\int_{\del}d^3 x\,\sqrt{\gamma}\,K,\label{full4d}
\end{align}
where the Gibbons-Hawking term is evaluated at the $AdS_4$ boundary.
This action admits the black hole solution eq.\eqref{RNBH} and the scalar field $\psi$ is charged under the background gauge field.

Varying the metric, we get the Einstein equations
\begin{align}
R_{\mu\nu}-\half g_{\mu\nu} R+\Lambda g_{\mu\nu}=16\pi G_N (T_{\mu\nu}^{\psi}+T_{\mu\nu}^F),\label{einsteineq}
\end{align}
where $T_{\mu\nu}^{\psi}$ and $T_{\mu\nu}^F$ are contributions to the stress tensor from the scalar field and gauge field respectively and are given by
\begin{align}
T_{\mu\nu}^F&=-\frac{1}{2G_N}\pqty{F_{\mu\rho}F_{\nu}^{\,\rho}-\quarter\,g_{\mu\nu}F^2},\label{fstress}\\
T_{\mu\nu}^{\psi}&=\frac{1}{G_N}\pqty{\half({D}_{\mu}\psi)^{\dagger}{D}_{\nu}\psi+\half({D}_{\nu}\psi)^{\dagger}{D}_{\mu}\psi-\half g_{\mu\nu}(g^{\alpha\beta}\,({D}_{\alpha}\psi)^{\dagger}{D}_{\beta}\psi+m^2\psi^{\dagger}\psi)}\label{mstress1}.
\end{align}
Note that when $\psi=0$ and $J_r=0$, we get the RN black hole soution given in eq.\eqref{RNAdS4II} which we will denote by $\bar{g}_{\mu\nu}$ below.
Varying the gauge field gives the equation of motion,
\begin{align}
\del_{\mu}(\sqrt{g}\,g^{\mu\rho}g^{\nu\sigma}F_{\rho\sigma})=e\,\sqrt{g}\,J^{\nu}\label{gaugeeom},
\end{align}
which can be solved to obtain
\begin{align}
F_{tr}=-\frac{Q}{\sqrt{g}\,g^{rr}g^{tt}}+e\,g_{tt}\,\del_t^{-1}J_r.\label{ftrsol}
\end{align}

Let us consider $S$-wave perturbations in the metric and work in the Fefferman-Graham gauge $\delta g_{\mu r}=0$,
\begin{align}
ds^2=a^2\,(1+h_{tt})\,dt^2+a^{-2}dr^2+b^2\,(1+h_{\theta\theta})\,d\Omega_2^2\label{metpert},
\end{align}
where $a^2$ and $b^2$ are given in eq.\eqref{RNAdS4II}. 
These metric fluctuations induce perturbations in the gauge field.
We have,
\begin{align}
F_{tr}&=\bar{F}_{tr}+F_{tr}^J+\delta F_{tr},\nonumber\\
\bar{F}_{tr}&=-\del_r \bar{A}_t=-\frac{Q}{\sqrt{\bar{g}}\,\bar{g}^{rr}\bar{g}^{tt}}, \label{gaugedef1} \\
\quad F_{tr}^J&=e\,\bar{g}_{tt}\,\del_t^{-1}J_r, \label{gaugedef2}\\
\quad \delta F_{tr}&=\del_t\delta A_r\label{gaugedef3},
\end{align}
where  $\bar{F}_{tr}$ is the background field strength, $F_{tr}^J$ is the response to the scalar current, upto the required order for the four-point function calculation,  and $\delta F_{tr}$ is the additional perturbation which arises due to metric fluctuations as we will see below. 
It is convenient to work in the gauge where $\bar{A}_t$ is the nonzero component for the background gauge field and the perturbations in the gauge field have non-zero component only along the $r$ direction. We denote the perturbation produced by the source $J$, eq.(\ref{gaugedef2}) as $A^J$, so that the full perturbation has a non-zero component $A^J_r+ \delta A_r$.

$\delta A_r$ can be obtained by linearising the first term in  eq.\eqref{ftrsol},
\begin{align}
&-\frac{Q}{\sqrt{g}\,g^{rr}g^{tt}}=-\frac{Q}{\sqrt{\bar{g}}\,\bar{g}^{rr}\bar{g}^{tt}}\pqty{1+\frac{1}{2}h_{tt}-h_{\theta\theta}}=\bar{F}_{tr}+\delta F_{tr}\nonumber\\
&\Rightarrow\delta A_r=-\bar{F}_{tr}\,\pqty{-\half \del_t^{-1}h_{tt}+\del_t^{-1}h_{\theta\theta}}\label{gaugepert}.
\end{align}
Expanding the action eq.\eqref{full4d} and using the equations of motion eqs.\eqref{einsteineq} and \eqref{gaugeeom}, we get the onshell action,
\begin{align}
S_{OS}&=\frac{1}{2}\int d^4x\,\sqrt{\bar{g}}\,\delta g^{\mu\nu}\,T_{\mu\nu}^{\psi}-\frac{e}{2G_N}\int d^4x\,\sqrt{\bar{g}}\,(A_{r}^J+\delta A_r)J^{r}\nonumber\\
&=-2\pi\int dt\, dr\,\pqty{\frac{b^2}{a^2}h_{tt}T_{tt}^{\psi}+2h_{\theta\theta} T_{\theta\theta}^{\psi}}-\frac{2\pi e}{G_N}\int dt\, dr\,b^2a^2(A_{r}^J+\delta A_r)J_{r}\label{onsh1}.
\end{align}
To get the four-point function for the field $\psi$, we express the perturbations in terms of the sources and compute the action to quartic order in $\psi$. 

In the discussion below we will systematically examine the terms appearing in eq.\eqref{onsh1}. 
We define,
\begin{align}
T_1&=-2\pi\int dt\, dr\,\pqty{\frac{b^2}{a^2}h_{tt}T_{tt}^{\psi}+2h_{\theta\theta}T_{\theta\theta}^{\psi}}\label{t1},\\
T_2&=-\frac{2\pi\,e}{G_N}\int dt\, dr\,b^2a^2\,A_{r}^JJ_{r}\label{t2},\\
T_3&=-\frac{2\pi\,e}{G_N}\int dt\, dr\,b^2a^2\,\delta A_{r}J_{r}\label{t3}.
\end{align}

\subsection*{Evaluating $T_1$}
Consider first the term $T_1$, eq.\eqref{t1}. This is infact similar to the  expression one would obtain in the four-point function computation of an uncharged scalar field with a background magnetic charge, see section 3 of \cite{Nayak:2018qej}. To simplify it, consider  the $tr$ and $rr$ equations in eq.\eqref{einsteineq},
\begin{align}
&\pqty{\frac{a'}{a}-\frac{b'}{b}}\del_t h_{\theta\theta}-\del_t\del_r h_{\theta\theta}=16\pi G_N (T_{tr}^{\psi}+T_{tr}^{F})=16\pi G_N (T_{tr}^{\psi}+T_{tr}^{J}),\label{treq}\\
&\frac{1}{a^4}\del_t^2h_{\theta\theta}+\pqty{\frac{a'}{a}+\frac{b'}{b}}\del_r h_{\theta\theta}+\frac{b'}{b}\del_r h_{tt}+\frac{1}{a^2b^2}h_{\theta\theta}=16\pi G_N (T_{rr}^{\psi}+T_{rr}^F)\nonumber\\
\Rightarrow&\frac{1}{a^4}\del_t^2h_{\theta\theta}+\pqty{\frac{a'}{a}+\frac{b'}{b}}\del_r h_{\theta\theta}+\frac{b'}{b}\del_r h_{tt}+\frac{1}{a^2b^2}\pqty{1-\frac{8\pi Q^2}{b^2}}h_{\theta\theta}=16\pi G_N (T_{rr}^{\psi}+T^J_{rr})\label{rreq},
\end{align}
where we have taken the terms involving the metric perturbation in $T_{\mu\nu}^F$ to the LHS. Here $T_{\mu\nu}^J$ is the $O(J)$ term in the stress tensor $T_{\mu\nu}^F$, eq.\eqref{fstress}, and is defined as
\be
\label{deftfj}
T^{J}_{\mu\nu}= -\frac{1}{2G_N}\,\pqty{\bar{F}_{\mu\rho}\,(F^J)_{\nu}^{\,\rho}+F^J_{\mu\rho}\,\bar{F}_{\nu}^{\,\rho}-\half \bar{g}_{\mu\nu}\,\bar{F}^{\alpha\beta}\,F^J_{\alpha\beta}},
\ee
where $\bar{F}_{\mu\nu}$ and $F^J_{\mu\nu}$ are given in eq.\eqref{gaugedef1} and eq.\eqref{gaugedef2} respectively and the raising and lowering of indices are done with the background metic $\bar{g}_{\mu\nu}$.
Also to the order we are interested in, we can write eq.\eqref{mstress1} as
\begin{align}
T_{\mu\nu}^{\psi}&=\frac{1}{G_N}\pqty{\half(\bar{D}_{\mu}\psi)^{\dagger}\bar{D}_{\nu}\psi+\half(\bar{D}_{\nu}\psi)^{\dagger}\bar{D}_{\mu}\psi-\half g_{\mu\nu}(g^{\alpha\beta}\,(\bar{D}_{\alpha}\psi)^{\dagger}\bar{D}_{\beta}\psi+m^2\psi^{\dagger}\psi)}\label{mstress},
\end{align}
where the raising and lowering of indices are with the background metric $\bar{g}_{\mu\nu}$.

Note that the combined contribution to the stress tensor coming from $\psi$, eq.\eqref{mstress}, and  $F^J_{\mu\nu}$, eq.\eqref{deftfj}, is conserved with respect to the background metric i.e.
\begin{equation}
\label{condscon}
\nabla_{[{\bar{g}}]}^{\mu} \pqty{T^{\psi}_{\mu\nu}+T^J_{\mu\nu}}=0,
\end{equation}
where $\nabla_{[\bar{g}]}$ denotes covariant derivative with respect to the background metric. 

To simplify further, we write the expression eq.\eqref{t1} as
\begin{align}
T_1&=-2\pi\int dt\, dr\,\pqty{\frac{b^2}{a^2}h_{tt}(T_{tt}^{\psi}+T_{tt}^J)+2h_{\theta\theta}(T_{\theta\theta}^{\psi}+T_{\theta\theta}^J)}\nonumber\\
&\qquad\qquad+2\pi\int dt\, dr\,\pqty{\frac{b^2}{a^2}h_{tt}T_{tt}^{J}+2h_{\theta\theta}T_{\theta\theta}^{J}}\label{t1exp}.
\end{align}
Let us denote the terms in the first and second lines of eq.\eqref{t1exp} as $T_{11}$ and $T_{12}$ respectively.
We can use the conservation equation eq.\eqref{condscon} and manipulate the equations, \eqref{treq} and \eqref{rreq}, analogous to \cite{Nayak:2018qej} and show that $T_{11}$ is given by 
\begin{align}
T_{11}&=-2\pi\int dt\, dr\,\pqty{\frac{b^2}{a^2}h_{tt}T_{tt}+2h_{\theta\theta}T_{\theta\theta}}\nonumber\\
&\qquad=-32\pi^2 G_N \int dt\, dr\, \left(\frac{2a^2b^3}{b'}\,T_{rr}\frac{1}{\del_t}T_{tr}-a^2b^2\Big(1+\frac{2a'b}{b'a}\Big)\,T_{tr}\frac{1}{\del_t^2}T_{tr}\right)\label{term11},
\end{align}
where
\begin{equation}
T_{\mu\nu}=T_{\mu\nu}^{\psi}+T_{\mu\nu}^J\label{fullstress}.
\end{equation}

Next, let us examine the contribution to the integral in $T_{11}$ from the near-horizon region. As discussed in section 4 of \cite{Nayak:2018qej}, we can use the conservation equations for $T_{\mu\nu}$ to express eq.\eqref{term11}  as a boundary term at the $AdS_2$ boundary,
\begin{align}
T_{11}&=-\frac{32\pi^2 G_N r_h^3}{L_2^2}\int_{bdy} dt\, (T_{tz}-z \partial_t T_{zz})\,\frac{1}{\del_t^4}\,(T_{tz}-z \partial_t T_{zz}),\label{term11bdy}
\end{align}
where we have kept the leading behaviour which is $O(r_h^3)$.
To simplify further, let us examine the behaviour of $T_{\mu\nu}^{\psi}$ and $T_{\mu\nu}^J$ at the $AdS_2$ boundary. From the form of $\psi$ at the $AdS_2$ boundary, eq.\eqref{asbeh}, we get for the leading non-contact terms, $\psi^\dagger \psi=\psi^\dagger \del_t\psi=\del_t\psi^\dagger\del_t\psi=0$. Therefore, from eq.\eqref{mstress},
\begin{align}
T_{tz}^{\psi}&=\frac{1}{2G_N} \pqty{\del_t\psi^{\dagger}\del_z\psi+\del_t\psi\del_z\psi^{\dagger}}-\frac{e}{2G_N}\bar{A}_t J_z=\tilde{T}_{tz}^{\psi}-e\,\bar{A}_t J_z,\label{ttzexp}\\
T_{zz}^{\psi}&=\frac{1}{G_N}\,\del_z\psi\del_z\psi^{\dagger}-\frac{1}{2G_N} \bar{g}_{zz}\pqty{\bar{g}^{\mu\nu}\del_{\mu}\psi^\dagger\del_{\nu}\psi+m^2\psi\psi^\dagger}=\tilde{T}_{zz}^{\psi}\label{tzzexp},
\end{align}
where we $\tilde{T}_{tz}^{\psi},\,\tilde{T}_{zz}^{\psi}$ are defined in eq.\eqref{deftildet}, $J_z$ is defined in eq.\eqref{jz} and $\bar{A}_t$ is gven in eq.\eqref{abart}.

Also, from eq.\eqref{deftfj}, we have
\begin{align}
T_{tz}^J&=0\label{tzj},\\
T_{zz}^J&=-\frac{1}{2G_N}\,\frac{eQ}{r_h^2}\,\frac{L_2^2}{z^2}\,\del_t^{-1}J_z\label{tzzj},
\end{align}
where we have used (see eqs.\eqref{gaugedef1}, \eqref{gaugedef2})
\begin{align}
\bar{F}_{tz}&=\frac{Q}{r_h^2}\,\frac{L_2^2}{z^2}\label{fbartz},\\
F_{tz}^J&=e\,\frac{L_2^2}{z^2}\,\del_t^{-1}J_z\label{fjtz}.
\end{align}
From eqs.\eqref{ttzexp}, \eqref{tzzexp}, \eqref{tzj}, \eqref{tzzj}, it is straightforward to verify that
\begin{equation}
\label{t11simp}
T_{tz}-z\del_t T_{zz}=\tilde{T}_{tz}^{\psi}-z\del_t \tilde{T}_{zz}^{\psi}.
\end{equation}
Therefore, eq.\eqref{term11bdy} becomes
\begin{align}
T_{11}&=-\frac{32\pi^2 G_N r_h^3}{L_2^2}\int_{bdy} dt\, (\tilde{T}_{tz}^{\psi}-z \partial_t \tilde{T}_{zz}^{\psi})\,\frac{1}{\del_t^4}\,(\tilde{T}_{tz}^{\psi}-z \partial_t \tilde{T}_{zz}^{\psi}).\label{t11final}
\end{align}
We see that the expression scales as $O(r_h^3)$ and agrees with eq.\eqref{Ttperm} of section \ref{bgf}.
The contribution of the region away from the horizon to the integral eq.\eqref{term11} gives an $O(r_h)^2$ contribution and is subleading.

Let us now examine the second line of eq.\eqref{t1exp}, $T_{12}$. From the expression for the stress tensor $T_{\mu\nu}^J$, eq.\eqref{deftfj}, we obtain
\begin{align}
T_{tt}^J&=-\frac{e}{2G_N}\,a^4\,\bar{F}_{tr}\del_t^{-1}J_r\label{tttj},\\
T_{\theta\theta}^J&=\frac{e}{2 G_N}\,b^2a^2\,\bar{F}_{tr}\del_t^{-1}J_r\label{tththj}.
\end{align}
Therefore, we have
\begin{align}
T_{12}&=2\pi\int dt\, dr\,\pqty{\frac{b^2}{a^2}h_{tt}T_{tt}^{J}+2h_{\theta\theta}T_{\theta\theta}^{J}}\nonumber\\
&=-\frac{2\pi e}{G_N}\int dt\,dr\,b^2a^2\,\bar{F}_{tr}\pqty{-\half\del_t^{-1}h_{tt}+\del_t^{-1}h_{\theta\theta}}J_r,\label{t12final}
\end{align}

We will show that the expression eq.\eqref{t12final} cancels with the term $T_3$, eq.\eqref{t3}.

\subsection*{Evaluating $T_{2} $ and $T_3$}
Let us now examine the remaining terms in the action, eqs.\eqref{t2} and \eqref{t3}. From eqs.\eqref{solar} and \eqref{gaugepert}, we get
\begin{align}
T_2=&-\frac{2\pi\,e^2}{G_N}\int dt\,dr\,b^2\,a^4\,J_r\,\del_t^{-2}J_r,\label{term2final}\\
T_3=&\frac{2\pi \,e}{G_N}\int dt\,dr\,b^2a^2\,\bar{F}_{tr}\,\pqty{-\half \del_t^{-1}h_{tt}+\del_t^{-1}h_{\theta\theta}}\,J_r\label{term3final}.
\end{align}

The expression eq.\eqref{term2final} is same as what we obtained for the probe $U(1)$ case as in section \ref{probeu1}, see eq.\eqref{gfos}. Therefore a similar calculation as done in appendix \ref{probe2du1} follows resulting in an $r_h^3$ enhanced contribution to the four-point function which is given by eq.\eqref{sj}.

Moreover, the expression eq.\eqref{term3final} cancels with the term $T_{12}$, eq.\eqref{t12final}. 

 Therefore, in the final expression for the onshell action eq.\eqref{onsh1}, the leading contribution which goes like $O(r_h^3)$ comes from the terms eq.\eqref{sbdypsi} and eq.\eqref{sj}. These terms can also be obtained by the exchange of the time reparamerisation and phase modes as explained in section \ref{bgf}.

\section{Phase Mode Action }
\label{2dbgf}
In appendix \ref{4dbgf}, we derived the onshell 4-D action for a scalar charged under the background gauge field. Here we derive a 2-D action by dimensional reduction and going to the near horizon $AdS_2$ region. We start with the 4-D action
\begin{align}
S=-\frac{1}{16 \pi G_N}\int d^4x\,\sqrt{g}\,\pqty{R-2\Lambda}-\frac{1}{4G_N}\int d^4x\,\sqrt{g}\,F^2-\frac{1}{8\pi G_N}\int_{\del}d^3 x\,\sqrt{\gamma}\,K\label{act4d}.
\end{align}
Let us now dimensionally reduce this action by taking the metric to be of the form
\begin{equation}
\label{dimmet}
ds^2=g_{\alpha\beta}(t,r)\,dx^{\alpha}dx^{\beta}+\Phi(t,r)^2\,d\Omega_2^2,
\end{equation}
where $\Phi$ is the dilaton field and $g_{\alpha\beta}$ is the 2-D metric in the $r-t$ plane.  Since we are interested in the near horizon region and to make contact with the JT model  \cite{Teitelboim:1983ux,JACKIW1985343,Almheiri:2014cka}, we do a Weyl rescaling of the metric,
\begin{equation}
\label{weyl}
g_{\alpha\beta}\rightarrow \frac{r_h}{\Phi}\,g_{\alpha\beta}.
\end{equation}
The action eq.\eqref{act4d} becomes
\begin{align}
S&=-\frac{1}{4G_N}\int d^2x\,\sqrt{g}\,\pqty{\frac{2r_h}{\Phi}+\Phi^2 R-2 r_h \Phi \Lambda}-\frac{\pi}{G_N}\int d^2x\,\sqrt{g}\,\frac{\Phi^3}{r_h}\,F_{\alpha\beta}F_{\gamma\delta}\,g^{\alpha\gamma}\,g^{\beta\delta}\nonumber\\
&\qquad -\frac{1}{2G_N}\int_{bdy}\sqrt{\gamma}\,\Phi^2 K\label{act2d}.
\end{align}
We are interested in this action in the near-$AdS_2$ region. The boundary term in eq.\eqref{act2d} is then evaluated at the boundary of this region

Let us expand the dilaton in a perturbation in $\phi$ about its  horizon value $r_h$,
\begin{equation}
\label{dilpert}
\Phi=r_h(1+\phi).
\end{equation}
We keep only terms upto $O(\phi)$ in the action; higher powers of $\phi$ do not contribute to the four-point function at $O(r_h^3)$ as argued in \cite{Nayak:2018qej}.

To $O(\phi^0)$, the action eq.\eqref{act2d} is
\begin{align}
S_0&=-\frac{r_h^2}{4G_N}\pqty{\int d^2x\,\sqrt{g}\,R+2\int_{bdy}\sqrt{\g}K}-\frac{4\pi Q^2}{G_Nr_h^2}\int\,d^2 x\,\sqrt{g}\nonumber\\
&\qquad\qquad-\frac{r_h^2}{G_N}\,\pi\int d^2x\,\sqrt{g}\,g^{\alpha\gamma}\,g^{\beta\delta}\,\pqty{F_{\alpha\beta}F_{\gamma\delta}-F_{\alpha\beta}^QF_{\gamma\delta}^Q},\label{act2dphi0}
\end{align}
where $F^Q_{\alpha\beta}$ is given by eq.\eqref{fqrt}.
The first two terms in eq.\eqref{act2dphi0} are topological. The third term in the first  line diverges but the divergence is cut off by corrections to the $AdS_2$ geometry. The correct expression for this term is obtained as follows. In the extremal or near extremal geometry, the action eq.\eqref{act4d} gives, on shell, the grand canonical partition function $S_{GC}$. This is related to the canonical ens emble partition function, obtained from $S_{CE}$ as
\begin{equation}
S_{GC}=S_{CE}-\frac{1}{2G_N}\int d^4x\,\sqrt{g_4}\,F^2,\label{sgc}
\end{equation}
where $g_4$ corresponds to the 4-D metric.
The canonical partition function $S_{CE}$ is  given by the topological piece at extremality. And  the extra piece in eq.\eqref{sgc} is the more correct version of  the second term, in eq.\eqref{act2dphi0}, accounting for corrections away from the near-horzion region. This gives 
\begin{align}
\frac{1}{2G_N}\int_{r_h}^{\infty}dr\int d^3x\,\sqrt{g_4}\,F^2=\frac{4\pi Q^2}{G_Nr_h}\int dt
\end{align}
where we used
\begin{equation}
\label{frt}
F_{rt}=\frac{Q}{\sqrt{g_4}\,g^{rr}g^{tt}}=\frac{Q}{r^2}.
\end{equation}

Therefore to $O(\phi^0)$, the action eq.\eqref{act2dphi0} can be written as
\begin{align}
S_0&=-\frac{r_h^2}{4G_N}\pqty{\int d^2x\,\sqrt{g}\,R+2\int_{bdy}\sqrt{\g}K}-\frac{4\pi Q^2}{G_Nr_h}\int dt\nonumber\\
&\qquad\qquad-\frac{r_h^2}{G_N}\,\pi\int d^2x\,\sqrt{g}\,g^{\alpha\gamma}\,g^{\beta\delta}\,\pqty{F_{\alpha\beta}F_{\gamma\delta}-F_{\alpha\beta}^QF_{\gamma\delta}^Q}.\label{act2dphi02}
\end{align}

The second term can be written as, 
\be
\label{actpb}
 -\frac{4\pi Q^2}{G_Nr_h}\int dt=-\pqty{\frac{r_h^2}{G_N}}\,4\pi\,{1\over r_h} \int dt \,\dot{\bar\theta}^2,
\ee
where 
\be
\label{defthetabar}
{\dot{\bar \theta}}=\frac{Q}{r_h}.
\ee
Eq.(\ref{actpb}) is of the same  form, upto a factor of $2$, as the action obtained for the phase mode in section \ref{probeu1}, eq.\eqref{actp2}.

Expanding eq.\eqref{act2d} to $O(\phi^1)$ gives
\begin{align}
S_1&=-\frac{r_h^2}{2G_N}\int d^2x\,\sqrt{g}\,\phi\, (R-\Lambda_2)-\frac{r_h^2}{G_N}\int_{bdy}\phi\,\sqrt{\g}\,K\nonumber\\
&\qquad\qquad -\frac{r_h^2}{G_N}\,3\pi\int d^2x\,\sqrt{g}\,\phi\,g^{\alpha\gamma}\,g^{\beta\delta}\,\pqty{F_{\alpha\beta}F_{\gamma\delta}-F_{\alpha\beta}^QF_{\gamma\delta}^Q}\label{act2dphi1}.
\end{align}
Here the terms in the first line correspond to the JT model, where $\Lambda_2$ is the two dimensional cosmological constant, $\Lambda_2=-\frac{2}{L_2^2}$.

Combining eq.\eqref{act2dphi0} and eq.\eqref{act2dphi1}, the complete action, including the boundary terms gives eq.\eqref{final2d}.
The terms in the last line of eq.\eqref{final2d} are the ones in JT action, including the last term which is a counter term needed for finiteness, see section 4 of \cite{Nayak:2018qej}. Here $F^Q$ is defined in eq.\eqref{fqrt}.

Let us make one comment before proceeding.  We have not been very explicit about the boundary condition to be satisfied by the gauge field perturbation above and similarly not been explicit about a possible boundary term involving the perturbation which may be needed in the action, eq.\eqref{act2d}. The explicit calculation we do carry out are done directly in $4$-D as discussed in section \ref{bgf}, the $2$-D action discussed here provides motivation for the action of the  phase mode and time reparametrisation mode  as we discuss further below.

Let us turn to the phase mode next and go back to the $4$ dimensional action, eq.\eqref{act4d} for analysing it. 
Consider a  gauge field configuration expanded about the background: 
\be
\label{solgf}
F_{rt}= \frac{Q}{\sqrt{g}\,g^{rr}g^{tt}} - \partial_t \delta A_r,
\ee
where 
\be
\label{defperta}
\delta A_r = {r_h\over \sqrt{g}\, g^{rr} g^{tt}} \delta \theta.
\ee
Here the perturbation is analogous to that in the probe gauge field case, eq.(\ref{formar}).

Plugging eq.(\ref{solgf}), eq.(\ref{defperta}) into eq.(\ref{act4d}) we get a term quadratic in $\delta \theta$, eq.\eqref{sphase}.
 We see that $\delta \theta$ behaves like the phase mode. 
 Note that the coefficient is suppressed by $\frac{1}{r_h}$ similar to the Schwarzian term above.

Next we analyse how this phase mode couples to time reparametrisations. 
In the asymptotically $AdS_2$ region, working in Poincare coordinates, eq.\eqref{zdef}, we get  from eq.(\ref{solgf})
\be
\label{valfz}
F_{zt}=-\frac{Q}{r_h^2}\,\frac{L_2^2}{z^2}-\del_t\delta A_z.
\ee
We work in the gauge where 
\begin{equation}
\label{gchoice}
A_\mu={\bar A}_\mu + \delta A_\mu
\end{equation}
where $\bar{A}_t$ is given in eq.\eqref{abart} and
\begin{align}
\delta A_z&=\frac{1}{r_h}\,\frac{L_2^2}{z^2}\,\delta \theta\label{delaz},
\end{align}
with the other components vanishing.

  To  find the coupling of this mode with the time reparametrisation mode it is convenient to change gauge.
A gauge transformation,
\begin{align}
\tilde{A}_{\alpha}=A_{\alpha}+\del_{\alpha}\pqty{\frac{1}{r_h}\,\frac{L_2^2}{z}\,\delta\theta(t)}\label{newA}.
\end{align}
allows us to go to the gauge in which the $z$ component of the gauge field is zero. We have 
\begin{equation}
\label{newaa}
\tilde{A}_z=0,\quad \tilde{A}_t=\frac{1}{r_h}\,\frac{L_2^2}{z}\pqty{\frac{Q}{r_h}+\delta\dot{\theta}}.
\end{equation}
We drop the tildes henceforth.

We think of the boundary time reparamtrisation as being embedded in the asymptotic isometry of $AdS_2$.
Under the transformation,
\begin{align}
t'&=t+\epsilon(t)\nonumber\\
z'&=z(1+\dot{\epsilon})\label{coord},
\end{align}
the gauge field transforms as
\begin{align}
A'_z(t',z')&=\frac{\del t}{\del z'}A_t=0,\nonumber\\
A'_t(t',z')&=\frac{\del t}{\del {t'}}A_t=\frac{1}{r_h}\frac{L_2^2}{z}\,\frac{1}{1+\dot{\epsilon}}\pqty{\frac{Q}{r_h}+\delta\dot{\theta}}\nonumber\\
\Rightarrow \frac{1}{r_h}\,\frac{L_2^2}{z'}\,\del_{t'}\theta(t')&=\frac{1}{r_h}\,\frac{L_2^2}{z}\,\frac{1}{1+\dot{\epsilon}}\pqty{\frac{Q}{r_h}+\delta\dot{\theta}}\nonumber\\
\Rightarrow \del_t\theta'&=\pqty{\frac{Q}{r_h}+\delta\dot{\theta}}(1+\dot{\epsilon})\label{thep}.
\end{align}
Therefore under the transformation eq.\eqref{coord},  $\delta\theta$ transforms to linear order in perturbations as
\begin{equation}
\label{delthtrans}
\delta\dot\theta\rightarrow \delta\dot\theta+\dot{\bar{\theta}}\,\dot\epsilon,
\end{equation}
where $\dot{\bar{\theta}}$ is given in eq.\eqref{defthetabar}.
This means that under the coordinate transformation eq.\eqref{coord}
the kinetic term for the phase mode transforms as
\begin{align}
S_{phase}&=-\frac{r_h^2}{G_N}\,2\pi\,\frac{1}{r_h}\int d\tau\,\pqty{\frac{d\delta\theta}{d\tau}}^2\rightarrow-\frac{r_h^2}{G_N}\,2\pi\,\frac{1}{r_h}\int d\tau\,\pqty{\frac{d\delta \theta}{d\tau}+\frac{d\epsilon}{d\tau}\,\dot{\bar{\theta}}}^2\label{finaltc}.
\end{align}

To calculate the four-point function, we couple the action eq.\eqref{final2d} to sources. The coupling between $\epsilon$ and $\psi$ comes from the kinetic part of the action for the field $\psi$, see appendix E of \cite{Nayak:2018qej} and we obtain eq.\eqref{scoup1}. The coupling between $\delta\theta$ and $\psi$ can be obtained in the same way as with the probe $U(1)$ in \ref{phase} and is given in eq.\eqref{scoup2}.
Combining all the contributions, we get the total 2-D action to be
\begin{align}
S_{2D}&=S_{mode}+S_{coup1}+S_{coup2},\label{s2dfin}
\end{align}
where the $S_{mode},\,S_{coup1},\,S_{coup2}$ are given in eqs.\eqref{modeact},\eqref{scoup1},\eqref{scoup2} respectively. Integrating out $\epsilon$ and $\delta\theta$, we get the final onshell action to be $S_{2D}=S_{bdy}^{\psi}+S_J$, see eqs.\eqref{sbdypsi},\eqref{sj}.

\bibliographystyle{JHEP}
\bibliography{refs}
\end{document}